\documentclass[twocolumn]{aastex701}

\usepackage{bm}
\usepackage{CJK}
\usepackage{amsmath}
\usepackage{graphicx}
\usepackage{booktabs}
\usepackage{tabularx}
\usepackage{etoolbox}
\makeatletter
\patchcmd{\frontmatter@title@produce}
  {\footnotetext{\hskip-11ptEmail: \@@email}}
  {\footnotetext{\hskip-11pt\parbox[t]{\dimexpr0.5\textwidth-0.5\columnsep\relax}{\raggedright Email: \@@email}}}
  {}{\typeout{WARNING: email footnote patch failed}}
\makeatother

\defcitealias{Ni_2025ApJ...995...96N}{Paper~I}

\begin{document}
\begin{CJK*}{UTF8}{gbsn}
\title{Radiation Hydrodynamics of Self-gravitating Protoplanetary Disks \\II. Accretion, Migration, Spin, and Internal Structure of GI Fragments}

\author[orcid=0000-0003-0794-1949]{Yang Ni (倪阳)}
\affiliation{Institute for Advanced Study, Tsinghua University, Beijing 100084, China}
\email[show]{ny22@mails.tsinghua.edu.cn}

\author[orcid=0000-0001-6858-1006]{Hongping Deng (邓洪平)}
\affiliation{Shanghai Astronomical Observatory, Chinese Academy of Sciences, Nandan Rd 80th, 200030 Shanghai, China}
\email[show]{hpdeng353@shao.ac.cn}

\author[0000-0001-6906-9549]{Xue-Ning Bai (白雪宁)}
\affiliation{Institute for Advanced Study, Tsinghua University, Beijing 100084, China}
\affiliation{Department of Astronomy, Tsinghua University, Beijing 100084, China}
\email[show]{xbai@tsinghua.edu.cn}

\begin{abstract}
In the first paper of this series, we showed that gravitational
instability (GI) in protoplanetary disks can produce fragments of
planetary rather than brown-dwarf mass, leaving the subsequent evolution
and ultimate fates of such fragments an open question. Here, we follow every
bound object in a global three-dimensional radiation hydrodynamic simulation
of a fragmenting $0.196\,M_\odot$ disk around a $1\,M_\odot$ star: seven
surviving fragments, one disrupted clump, and one merged clump, followed for
$1.2$\,kyr. The
fragments form at $1.4$--$3.0\,M_\mathrm{J}$, consistent with the
normalized initial-mass distribution of Paper~I. Mass growth follows a single
Hill-limited scaling $\propto \Sigma \Omega R_\mathrm{H}^2$, regulated by the
delivery of gas into draining feeding zones, while the latest-forming
fragments experience supply starvation. Migration is bidirectional and
governed by gravitational interactions with both the disk and neighboring
clumps. Accretion supplies most of the fragments' spin, which aligns with
their orbits and provides roughly a tenth of the support against gravity.
The interiors are entropy-stratified, largely convectively stable, and
accumulate mass far faster than they can radiatively cool: compression is
quasi-adiabatic, and central entropies near $12\,k_\mathrm{B}$ per baryon favor
hot-start initial conditions for fragments that evolve into gas giants.
Integrating the measured growth law against the measured gas budget, we predict
that the disk produces gas giants, brown dwarfs, and a possible low-mass
stellar companion, while one scattered fragment may become a free-floating
planet: the outcome of disk fragmentation is determined not merely at birth,
but by the subsequent mass supply and dynamics.
\end{abstract}


\section{Introduction}\label{sec:intro}

Gravitational instability (GI) is one of the two principal pathways
proposed for giant planet formation
\citep{Cameron_1978M&P....18....5C, Boss_1997Sci...276.1836B,
Kratter_2016ARA&A..54..271K,Deng_2021NatAs...5..440D}, complementary to core accretion
\citep{Pollack_1996Icar..124...62P, Lambrechts_2012A&A...544A..32L}
at wide separations and around low-mass stars, where the latter
struggles. The long-standing objection to GI is one of scale: disk
fragmentation was generally expected to produce overmassive clumps,
closer to brown dwarfs than to gas giants
\citep{Forgan2011jeans, Kratter_2016ARA&A..54..271K,
Xu_2025ApJ...986...91X}. In the first paper of this series
\citep[][hereafter Paper~I]{Ni_2025ApJ...995...96N}, we addressed this
objection with global radiation hydrodynamic (RHD) simulations that
resolve the local collapse. We found that fragments are born at
planetary masses, distributed around a universal mass scale set by the
local conditions in the fragmenting spiral arms. These planetary birth masses, however, do not by themselves determine the
fragments' eventual fates.
As a fragment accretes and contracts within its parent disk, its
changing mass and size modify the gravitational torques driving its orbital
evolution
\citep{Zhu_2012ApJ...746..110Z}.
This orbital evolution in turn changes both the gas available for
further accretion and the tidal forces that the fragment's evolving
structure must withstand
\citep{Boley_2010Icar..207..509B,2012MNRAS.427.1725G}.
How this coupled evolution determines the eventual fates of initially
planetary-mass fragments remains an open question.

Directly constraining this post-formation evolution with observations, however, has proven exceptionally difficult. Gravitational instability is expected to be most prevalent during the earliest, most massive Class~0/I stages of star formation, when the protoplanetary disk is still deeply embedded within an infalling protostellar envelope and obscured by high dust optical depths \citep{Kratter_2016ARA&A..54..271K}. Consequently, identifying infant planetary-mass fragments, tracing their localized circumplanetary feeding flows, or directly measuring their migration remains challenging even with current high-contrast imaging and interferometry. Candidate forming protoplanets embedded in self-gravitating spiral arms—most notably AB~Aurigae~b \citep{Currie_2022NatAs...6..751C,Speedie_2024Natur.633...58S}—provide tantalizing evidence for wide-orbit formation, yet such systems remain rare, and existing observations do not yet establish the coupled dynamical and structural evolution of individual fragments.

Theoretical modeling therefore provides the principal framework for investigating this post-fragmentation evolution; yet (semi-)analytical treatments and population-synthesis models struggle to capture the inherently non-linear aftermath of disk fragmentation \citep{Kratter_2016ARA&A..54..271K}. The subsequent fate of a fragment is governed by a complex web of coupled mechanisms: the clump exchanges mass and angular momentum with the gravito-turbulent disk, experiences non-axisymmetric gravitational torques from spiral arms, and undergoes mutual gravitational encounters with neighboring clumps. Because these hydrodynamic and gravitational interactions operate across disparate spatial and temporal scales, analytical formulations must rely on idealized prescriptions for gas capture, migration rates, and tidal survival. Such approximations introduce substantial uncertainties, leaving the predicted survival rates and final mass distributions of GI fragments sensitive to the adopted prescriptions and parameter choices \citep{Forgan_2018MNRAS.474.5036F,Schib_2025A&A...704A..28S}.

Direct numerical simulations are therefore indispensable for capturing this non-linear evolution, and foundational studies have provided critical insights into fragment dynamics and growth \citep[e.g.,][]{Boley_2010Icar..207..509B,Baruteau_2011MNRAS.416.1971B,Zhu_2012ApJ...746..110Z}. In particular, two-dimensional simulations with self-consistent fragmentation demonstrated that clumps can accrete at rates scaling with their Hill sphere while migrating rapidly inward \citep{Zhu_2012ApJ...746..110Z}. However, two-dimensional calculations cannot resolve vertical gas circulation or the three-dimensional structure of a fragment, and radiative heating and cooling must be represented through vertically integrated or parameterized treatments. Furthermore, to circumvent the severe numerical expense of tracking local collapse within global disks, several three-dimensional studies represent embedded protoplanets with accreting sink particles \citep{Stamatellos_2015ApJ...810L..11S,Fletcher_2019MNRAS.486.4398F}. While sink particles permit longer integration times, they replace gas within an adopted accretion radius, leaving the fragment's internal contraction, redistribution of angular momentum, and response to tides on smaller scales unresolved. Three-dimensional RHD simulations have also followed both the orbital and thermal evolution of self-consistently formed fragments \citep{Tsukamoto_2013MNRAS.436.1667T}. However, those calculations employed flux-limited diffusion (FLD), which directs the radiative flux down the local radiation-energy gradient, limiting its accuracy for anisotropic transport in optically thin regions \citep{Kanno_2013PASJ...65...72K}. The resulting uncertainty in radiative cooling can propagate into both fragment contraction and the disk conditions that govern gas supply and migration.

Addressing these limitations requires global, three-dimensional simulations that resolve fragment interiors and multi-body interactions without sink particles, while treating radiation transport across both optically thick and thin regimes. In this second paper of our series, we carry out such an investigation using a global three-dimensional radiation hydrodynamic (RHD) simulation of a $0.196\,M_\odot$ self-gravitating disk around a $1\,M_\odot$ star, following the system for $\approx 1.2$\,kyr after the onset of fragmentation. Employing the M1 radiation closure developed in \citetalias{Ni_2025ApJ...995...96N}, the calculation accurately tracks anisotropic cooling without directional bias. We follow all nine self-consistently formed bound clumps—seven surviving fragments, one disrupted clump, and one merger—and construct per-object mass and angular-momentum budgets across their simulated histories. Using these budgets, we quantify how multi-fragment competition regulates gas delivery into local feeding zones, how torques from ambient gas and neighboring clumps govern orbital migration and dynamical scattering, and how ongoing accretion shapes internal entropy stratification, spin alignment, and tidal resilience. Finally, by integrating the measured growth rates against the parent disk's remaining gas budget, we assess the conditional long-term fates of the fragments and delineate the diverse spectrum of planetary, brown-dwarf, and stellar-mass outcomes produced by gravitational instability.

Section~\ref{sec:method} describes the numerical setup and analysis.
Section~\ref{sec:results} presents the fragment evolution and conditional
estimates of later masses. Section~\ref{sec:discussion} discusses the
implications for fragment fates and the limitations of the calculation,
and Section~\ref{sec:conclusions} summarizes our conclusions.

\section{Methods}\label{sec:method}

\subsection{Numerical method}\label{sec:numerics}
Our simulation is performed with the meshless finite-mass method in the
{\sc gizmo} code \citep{Hopkins_2015MNRAS.450...53H}, coupled to the
radiation-hydrodynamics module that we developed and tested in
\citetalias{Ni_2025ApJ...995...96N}. The module solves the
frequency-integrated radiation moment equations under the M1 closure
\citep{Levermore_1984JQSRT..31..149L} with an implicit--explicit scheme. Radiation
transport is advanced explicitly with a second-order Harten--Lax--van~Leer Riemann
solver and piecewise-linear reconstruction, and the stiff radiation--matter coupling
terms are integrated implicitly. A geometric outflow boundary condition allows thermal
radiation to escape from the disk surface. The governing equations, solver design, and an
extensive test suite are described in \citetalias{Ni_2025ApJ...995...96N}, and we do not
repeat them here.

The present run differs from the simulation suite of
\citetalias{Ni_2025ApJ...995...96N} in the disk model and run setup (see Section~\ref{sec:diskmodel}).
First, the gas equation of state uses an adiabatic index $\gamma = 1.4$ with a
mean molecular weight $\mu = 2.33$, appropriate for the cold, molecular
(H$_2$-dominated) gas of the outer disk, instead of the monatomic
$\gamma = 5/3$ adopted in \citetalias{Ni_2025ApJ...995...96N} for comparability with
earlier work. We note that our simulations are not run long enough to trace the second core formation and H$_2$ dissociation \citep[e.g.,][]{2012MNRAS.427.1725G}. In fact, even the fragment centres, the densest
regions we track, remain below $\sim\!10^3\,$K (see Section \ref{sec:interior}). Second, the gray power-law opacity of \citetalias{Ni_2025ApJ...995...96N} is replaced by
more realistic tabulated opacities of \citet{Zhu_2021MNRAS.508..453Z}, which assume a solar metallicity contributed by dust whose  size follows a power-law distribution with a slope $q = 3.5$
and a maximum grain size $a_\mathrm{max} = 0.1\,\mathrm{cm}$. The table provides
the Rosseland and Planck mean opacities over
$10^{-14} \leq \rho/(\mathrm{g\,cm^{-3}}) \leq 0.79$ and
$1 \leq T/\mathrm{K} \leq 7.9\times10^{6}$.
Third, here the speed of light is reduced by a factor $\tilde{c}/c = 5\times10^{-4}$,
more conservative than that in \citetalias{Ni_2025ApJ...995...96N}, so that it
satisfies the static-diffusion criterion, even though $\tilde{c}/c = 4\times10^{-4}$ is
shown to be sufficient in the explicit tests of \citetalias{Ni_2025ApJ...995...96N}.
The particle mass of our run, $1.0\times10^{-4}\,M_\mathrm{J}$, is similar to that of
the \citetalias{Ni_2025ApJ...995...96N} simulations (Table~\ref{tab:setup}).\footnote{We
refer to the resolution elements of the meshless finite-mass method as particles. They
are called cells in \citetalias{Ni_2025ApJ...995...96N}.} We do not repeat a resolution
study here. The convergence test of \citetalias{Ni_2025ApJ...995...96N} (their
Appendix~B) shows that the mass history of the most massive clump is nearly unchanged
at quadruple resolution, which indicates that the accretion of well-resolved clumps is
converged at this resolution.

\subsection{Disk model}\label{sec:diskmodel}

We simulate a single massive disk of mass $M_\mathrm{d} = 0.196\,M_\odot$ around a
$1\,M_\odot$ star, initially extending from approximately $60$ to $180$\,au, with a surface
density profile $\Sigma \propto R^{-7/4}$ and a temperature profile
$T \propto R^{-1/2}$, normalized to $75$\,K at $100$\,au. These slopes
place the disk at a uniform Toomre parameter
$Q = c_s\kappa/(\pi G \Sigma) = 2.0$ across its full radial extent, where
$c_s = (k_\mathrm{B}T/\mu m_\mathrm{H})^{1/2}$ is the isothermal sound
speed and $\kappa$ is the epicyclic frequency. The disk is resolved
by $2{,}038{,}732$ equal-mass particles of $9.63\times10^{-8}\,M_\odot$
($1.0\times10^{-4}\,M_\mathrm{J}$) each. The star is treated as an external point-mass
potential. From this configuration, the disk cools, develops rings and spiral
arms, and forms its first bound clump at $t = 2.46\,\mathrm{kyr}$. We run the simulation
for a total of $3.67\,\mathrm{kyr}$ (hereafter $t_\mathrm{end}$), i.e., $\approx 3.7$ orbits
at $100$\,au, so the fragment population is followed for $\approx 1.2$\,kyr after the onset
of fragmentation.
Gas that spreads beyond the radial boundaries of the computational domain, at $30$ and
$300$\,au, is removed. Over the course of the run, this removes $1.7$\% of the disk mass
($3.6\,M_\mathrm{J}$), with just over half lost through the inner boundary and the remainder
through the outer boundary; all of it is far from the fragments' orbits.
Simulation snapshots are taken every $24.15\,\mathrm{yr}$, and all Lagrangian analyses
below exploit the unique particle identities across snapshots. Table~\ref{tab:setup}
summarizes the setup alongside that of \citetalias{Ni_2025ApJ...995...96N}.

\begin{deluxetable}{lcc}
\tablecaption{Simulation setup of this work compared with the suite of
\citetalias{Ni_2025ApJ...995...96N}\label{tab:setup}}
\tablehead{
\colhead{Parameter} &
\colhead{\citetalias{Ni_2025ApJ...995...96N}} &
\colhead{This work}
}
\startdata
Star mass ($M_\odot$) & $1$ & $1$ \\
Disk mass ($M_\odot$) & $0.13$--$0.20$ & $0.196$ \\
Radial extent (au) & $30$--$100$ & $60$--$180$ \\
$\Sigma$ profile & $\propto R^{-1}$ & $\propto R^{-7/4}$ \\
$T$ profile & $\propto R^{-1/2}$ & $\propto R^{-1/2}$ \\
Initial Toomre $Q$\tablenotemark{a} & $1.5$ at $100$\,au & $2.0$ (uniform) \\
Adiabatic index $\gamma$ & $5/3$ & $1.4$ \\
Mean molecular weight $\mu$ & $2.33$ & $2.33$ \\
Opacity & $\min[(T/100\,\mathrm{K})^{2},1]\,\kappa_0$, $\kappa_0 = 1$ or $5\,\mathrm{cm^2\,g^{-1}}$ & tabulated\tablenotemark{b} \\
$\tilde{c}/c$ & $4\times10^{-4}$ & $5\times10^{-4}$ \\
Number of particles & $\sim 2\times10^{6}$ & $2.04\times10^{6}$ \\
Particle mass ($M_\mathrm{J}$) & $\sim 10^{-4}$ & $1.0\times10^{-4}$ \\
Output cadence (yr) & $10$ & $24.15$ \\
Duration (kyr) & $\leq 5$ & $3.67$ \\
\enddata
\tablenotetext{a}{Sound-speed conventions differ with the equations of state.
This work evaluates $Q$ with the isothermal sound speed
(Section~\ref{sec:diskmodel}).}
\tablenotetext{b}{Rosseland and Planck mean opacities from
\citet{Zhu_2021MNRAS.508..453Z}, described in Section~\ref{sec:numerics}.}
\end{deluxetable}

\subsection{Clump identification, tracking, and the fragment samples}\label{sec:clumpid}
\label{sec:clumpdef}

We identify bound structures, which we call \emph{clumps}, in every snapshot using
\texttt{CloudPhinder}\footnote{\url{https://github.com/mikegrudic/CloudPhinder}},
the structure finder used in \citetalias{Ni_2025ApJ...995...96N}. Starting from local
density peaks, it grows candidate structures particle by particle in order of
decreasing density and constructs, around each peak, the largest structure that is
energetically bound,
\begin{equation}\label{eq:bound}
    E_\mathrm{K} + U < |E_\mathrm{P}|,
\end{equation}
where $E_\mathrm{K}$, $U$, and $E_\mathrm{P}$ are the kinetic energy
(computed from velocity fluctuations about the structure's center-of-mass velocity), the internal energy,
and the gravitational self-energy of the structure.  We
apply the finder to particles above a density threshold of
$5.9\times10^{-14}\,\mathrm{g\,cm^{-3}}$. For the census, we additionally require that the
structure persists for at least two consecutive snapshots, which excludes
a transient bound over-density dispersed quickly. Every clump in the census contains
more than $1.3\times10^{4}$ particles in every snapshot.

To measure how far a clump has proceeded toward virial equilibrium, we use the
virial parameter
\begin{equation}\label{eq:alphavir}
    \alpha_\mathrm{vir}
    = \frac{2E_\mathrm{K} + 3(\gamma-1)\,U}{|E_\mathrm{P}|}.
\end{equation}
This definition generalizes the virial parameter in
\citetalias{Ni_2025ApJ...995...96N} by explicitly accounting for different ratios of specific heats.
A self-gravitating ideal gas in virial equilibrium satisfies
$2E_\mathrm{K} + 3(\gamma-1)U = |E_\mathrm{P}|$, where the pressure integral contribution is
$3\int P\,\mathrm{d}V = 3(\gamma-1)U$, so that $\alpha_\mathrm{vir} = 1$. We note that $\alpha_\mathrm{vir} = 2$ is equivalent to the marginally bound criterion in Equation~(\ref{eq:bound}) when $\gamma=5/3$ (\citetalias{Ni_2025ApJ...995...96N}).

We follow each clump through the identities of its member particles
\citep[as in][]{Ni_2025A&A...699A.282N}. A clump is linked to every clump in the next
snapshot that shares its particles, and the sequence of links that carries the largest
fraction of its mass forms its track. A track begins at the snapshot in which the clump
is first identified and ends when the run ends, when the clump is disrupted, or when it
merges with another clump. A clump is disrupted when no clump in the next snapshot
contains any of its particles. The run yields nine tracks, one of which ends in a
disruption and one in a merger (Section~\ref{sec:census}). Following
\citetalias{Ni_2025ApJ...995...96N} we define the \emph{identification epoch}
$t_\mathrm{identify}$ of a clump as the first snapshot at which it becomes marginally gravitationally bound. Its \emph{initial mass}
$m_\mathrm{frag}$ is the bound mass at that epoch, and its initial
orbital radius $R_\mathrm{frag}$ is the cylindrical radius of its
center of mass at the same epoch. We measure its size by the effective
radius
$R_\mathrm{eff} = [\,(5/3)\textstyle\sum_i m_i |\bm{r}_i - \bm{r}_c|^2 /
\sum_i m_i\,]^{1/2}$, where $m_i$ and $\bm{r}_i$ are the mass and position of member
particle $i$ and $\bm{r}_c$ is their center of mass. $R_\mathrm{eff}$ is the radius of a
uniform-density sphere with the same mass-weighted mean-square radius, as in
Equation~(36) of \citetalias{Ni_2025ApJ...995...96N}.

We define \emph{fragments} as the clumps that survive, without disruption,
to the end of the simulation. Seven fragments form during the simulation, labeled F1--F7 in order of
formation time. We note that F6 and F7 are first identified in the same snapshot and distinguished by
their initial masses. The disrupted clump is labeled D1, and the late-forming
clump that merges into F1 is labeled M1.

\subsection{Analysis conventions}\label{sec:analysis}

After its formation, a fragment continues to interact with the surrounding disk and
with the other clumps. As Section~\ref{sec:results} shows, it accretes gas from its
surroundings, migrates under the torques from the disk gas and the other clumps, gains
spin from the gas it accretes, and contracts and heats as its mass grows. To quantify
these processes, we define below the local disk environment of a fragment and the
diagnostics of its accretion, growth episodes, torques, spin, and interior structure.
We also define a matched age, which allows us to compare fragments that form at
different times.

\emph{Ambient gas and local disk quantities.} The \emph{ambient} gas at a
snapshot is all gas belonging to no cataloged clump. Local disk
quantities around a fragment are computed from ambient gas only, so that
a fragment's own envelope and its neighbors' envelopes do not contaminate
its environment. A fragment's orbital radius $a$ is the cylindrical
radius of its center of mass, and orbital frequencies are taken
Keplerian, $\Omega = (GM_\star/a^{3})^{1/2}$, as an ad hoc approximation. We note that the measured rotation of
the ambient gas deviates from the Keplerian rate by several percent, and
locally more within arms, through the disk's self-gravity and pressure
gradients, but the quantities built from $\Omega$, namely the rate scale
$\Sigma_\mathrm{loc}\Omega R_\mathrm{H}^{2}$, the scale height $H$, the
linear migration timescale, and the analytic
initial-mass scale of Section~\ref{sec:initialmass}, are used either with
measured normalizations or in comparisons at the factor-of-a-few level,
so the deviation shifts no conclusion. The
orbital eccentricity $e$ is the two-body osculating value computed from
the center-of-mass position $\bm{r}$ and velocity $\bm{v}$ about the star
alone,
\begin{equation}\label{eq:ecc}
    e = \left|\,
    \frac{\bm{v} \times (\bm{r} \times \bm{v})}{G M_\star}
    - \frac{\bm{r}}{|\bm{r}|}\,\right| .
\end{equation} The Hill radius is
$R_\mathrm{H} = a\,[m/(3M_\star)]^{1/3}$, with $m$ the fragment's bound
mass. The local surface density $\Sigma_\mathrm{loc}$ is the ambient mass
within the in-plane circle of radius $2R_\mathrm{H}$ around the fragment,
divided by the area of that circle. The local sound speed $c_s$ is the
mass-weighted isothermal sound speed of the same gas, the local scale
height is $H = c_s/\Omega$, and the thermal (Bondi) radius is
$R_\mathrm{B} = Gm/c_s^{2}$. The \emph{feeding annulus} of a fragment is the part of the disk at cylindrical radii
within $2R_\mathrm{H}$ of its orbital radius $a$. The ambient mass in this annulus is
\begin{equation}\label{eq:mfeed}
    M_\mathrm{feed} = \sum_{|R_i - a| < 2R_\mathrm{H}} m_i ,
\end{equation}
where the sum runs over ambient particles, and $m_i$ and $R_i$ are the mass and
cylindrical radius of particle $i$. Section~\ref{sec:accretion} shows that this
extent, $2R_\mathrm{H}$ on each side of the orbit, matches the measured
origins of the gas the fed fragments F1--F4 accrete.

\emph{Mass accretion.} Two quantities enter the mass budget: the
Hill inventory $M(<R_\mathrm{H})$, which is the total gas mass inside the
sphere of radius $R_\mathrm{H}$ centered on the fragment, and its unbound
part $M_\mathrm{res} = M(<R_\mathrm{H}) - m$. The \emph{total inflow} over a stated interval is the mass
that crosses the Hill surface inward during it, counted without
subtracting the mass that crosses outward. The growth rate $\dot m$ used
throughout is the net rate of change of the bound mass, smoothed with a
centered running mean of width equal to the orbital shear time
$\Omega^{-1}$, evaluated at each fragment's median orbital radius. Growth rates are compared with the Hill rate through
the efficiency
$C_\mathrm{eff} \equiv \dot m/(\Sigma_\mathrm{loc}\Omega R_\mathrm{H}^{2})$,
whose denominator is smoothed over the same window as $\dot m$.
We discard intervals in which more than $5$\% (mass) of the gas in a
fragment's Hill sphere lies simultaneously inside a neighboring clump's
Hill sphere and, for F1, the
intervals spanning the merger. For the efficiency, discarded intervals are masked in both numerator
and denominator before the smoothing, and efficiency statistics use the
intervals with positive smoothed rate. We quote the median of
$C_\mathrm{eff}$ for each fragment alone and the median over the pooled
intervals of all seven.

\emph{Growth episodes.} Alongside the smooth growth, we identify growth episodes, which are short intervals
in which the fragment mass changes much faster than smooth accretion alone would
produce. Section~\ref{sec:accretion} describes the three episodes whose mechanism we
can identify.

\emph{Torques.} The gravitational torque on a fragment is measured from
the gravitational accelerations of individual particles in the simulation output: with $\bm{g}_i$ the total gravitational
acceleration on member particle $i$,
\begin{equation}\label{eq:torque}
    \Gamma_\mathrm{ext} = \sum_{i\,\in\,\mathrm{members}}
    m_i\,(\bm{r}_i \times \bm{g}_i)_z ,
\end{equation}
in which the star, whose force on every member is central, and the
members' mutual self-gravity, which cancels pairwise, contribute
exactly zero. The sum
is therefore the torque exerted by all matter outside the fragment.
We split it exactly by calculating the gravitational acceleration of
every member particle from two separate source sets: the member particles of every
other cataloged clump, which gives $\Gamma_\mathrm{clumps}$, and the
ambient gas, which gives
$\Gamma_\mathrm{gas}$. The hydrodynamic torque
$\sum m_i(\bm{r}_i \times \bm{a}_{\mathrm{hyd},i})_z$, evaluated from the
particles' hydrodynamic accelerations, enters the angular-momentum budget of
Section~\ref{sec:migration}. The radial drift rate $\dot a$ is the change of $a$ between consecutive
snapshots divided by the output interval. The drift timescale is
the per-interval $|a/\dot a|$, and comparisons with the linear
timescale quote the median of the per-interval ratio over each
fragment's life. The linear
(Type-I) reference timescale is
$t_\mathrm{I} = (1/2.7)\,(M_\star/m)\,(M_\star/\Sigma a^{2})\,(H/a)^{2}\,
\Omega^{-1}$ \citep[the constant-$\Sigma$ coefficient of][]{2002ApJ...565.1257T}, evaluated from the local ambient
disk, with $\Sigma$ taken from the feeding annulus and $H$ as above.

\emph{Spin.} The spin of a fragment is the angular momentum of its bound
members about their center of mass, in the center-of-mass velocity frame,
\begin{equation}\label{eq:spin}
    \bm{S} = \sum_{i\,\in\,\mathrm{members}} m_i\,
    (\bm{r}_i - \bm{r}_c) \times (\bm{v}_i - \bm{v}_c).
\end{equation}
Here $\bm{r}_c$ and $\bm{v}_c$ are the mass-weighted mean position and
velocity of the members. To separate ordered rotation from random
motions, we divide the members into twelve cylindrical shells of equal particle number about the spin axis $\hat{\bm{S}} = \bm{S}/|\bm{S}|$ and
define the rotational energy as the kinetic energy of the mean azimuthal
motion of those shells,
\begin{equation}\label{eq:erot}
    E_\mathrm{rot} = \sum_{k} \tfrac{1}{2}\,M_k\,
    \langle v_\phi\rangle_k^{2},
\end{equation}
where $M_k$ is the mass of shell $k$ and $\langle v_\phi\rangle_k$ the
mass-weighted mean azimuthal velocity about $\hat{\bm{S}}$ within it.
The rotational support is then
$\beta_\mathrm{spin} = E_\mathrm{rot}/|E_\mathrm{P}|$. The spin--orbit
obliquity is the angle between $\bm{S}$ and the orbital angular momentum
$m\,\bm{r}_c \times \bm{v}_c$. The degree of differential rotation
is measured by the power-law index $q_\Omega =
\mathrm{d}\ln\Omega_\mathrm{spin}/\mathrm{d}\ln\varpi$ of the shell
rotation profile $\Omega_\mathrm{spin}(\varpi) =
\langle v_\phi\rangle(\varpi)/\varpi$, with $\varpi$ the cylindrical
distance from the spin axis. The index is fitted by least squares in
the log--log plane over the shells whose mean radii lie between the
radii enclosing one quarter and one half of the bound mass. The spin delivered by accretion, $\Delta J_\mathrm{adv}$, is the
component of the spin angular momentum along the fixed rotation axis of
the disk carried by joining particles minus that carried by leaving particles,
evaluated in each epoch's center-of-mass frame. We report its running
total as a fraction of the change of the spin component along the same
fixed axis, $\Delta J_\mathrm{spin}$, accumulated to the same time.

\emph{Interior profiles.} Entropy is traced by the adiabat label
\begin{equation}\label{eq:adiabat}
    K \equiv (\gamma-1)\,u\,\rho^{1-\gamma},
\end{equation}
where $u$ is the specific internal energy of a particle, so that $U = \sum_i
m_i u_i$ over the bound members. The label is proportional to
$P/\rho^{\gamma}$ and is therefore constant for
any parcel of ideal gas that evolves adiabatically. A rise in $K$
measures heating beyond what compression alone provides, and a fall
measures cooling.

Because the gas obeys a fixed-$\gamma$ equation of state, $K$ also fixes
its specific entropy. With $c_V = k_\mathrm{B}/(\gamma-1)$ per molecule or atom, the
entropy per baryon is
\begin{equation}\label{eq:entropy}
    s = \frac{k_\mathrm{B}}{\mu\,(\gamma-1)}\,\ln\frac{K}{K_0} + s_0,
\end{equation}
with $1/[\mu(\gamma-1)] = 1.073$ for our gas, so that only the constant $s_0$
is undetermined. We fix it on the ideal-gas expression of
\citet{Marleau_2014MNRAS.437.1378M}, which gives $s_0 = 9.6\,k_\mathrm{B}$ per baryon at
$K_0 = 2.9\times10^{12}\,\mathrm{erg\,g^{-1}(g\,cm^{-3})^{-2/5}}$, the
value of $K$ at $1600$\,K and $3$\,bar. Our entropies then carry the zero point
that \citet{Marleau_2014MNRAS.437.1378M} use for the initial entropies of
young giant planets. In the following analysis, we quote the central value, defined with the other central quantities below.

Interior profiles are constructed at every output
snapshot of a fragment's life, about its density peak, defined as the
density-weighted centroid of its densest $1$\% of particles. Profiles, in radius or in
mass coordinate, are medians in $32$ shells of equal particle number, which is
equal mass since the particles have equal masses. In
interior contexts the half-mass radius
$R_\mathrm{half}$ is the median member radius about that density peak.
Central values are medians over the same densest $1$\%. Convective
stability is judged by the Schwarzschild criterion for an adiabatic
stratification, $\mathrm{d}K/\mathrm{d}r \geq 0$. The free-fall time is
$t_\mathrm{ff} = [3\pi/(32 G\rho_c)]^{1/2}$ at the central density
$\rho_c$, the accretion time is $t_\mathrm{acc} = m/\langle\dot m\rangle$ with
$\langle\dot m\rangle$ the lifetime-mean growth rate, and
$L = 4\pi R_\mathrm{half}^{2}\langle
F_r\rangle$ is the luminosity obtained from the mean outward radiative
flux $\langle F_r \rangle$ of the particles whose radii lie within ten
percent of the half-mass radius. The
Kelvin--Helmholtz time is $t_\mathrm{KH} = |E_\mathrm{P}|/L$, the time in
which a fragment could radiate its binding energy away at that
luminosity. The accretion power is
$L_\mathrm{acc} = Gm\langle\dot m\rangle/R_\mathrm{half}$.

\emph{Matched age.}
Where fragments are compared at a common age we use
the first $434.7$\,yr after each fragment's identification ($18$ output
intervals). This is the age of the youngest fragments at the end of the
run, and therefore the longest span over which all seven can be
compared.

\section{Results}\label{sec:results}

\subsection{Fragment census and fates}\label{sec:census}

\begin{figure*}[ht!]
\plotone{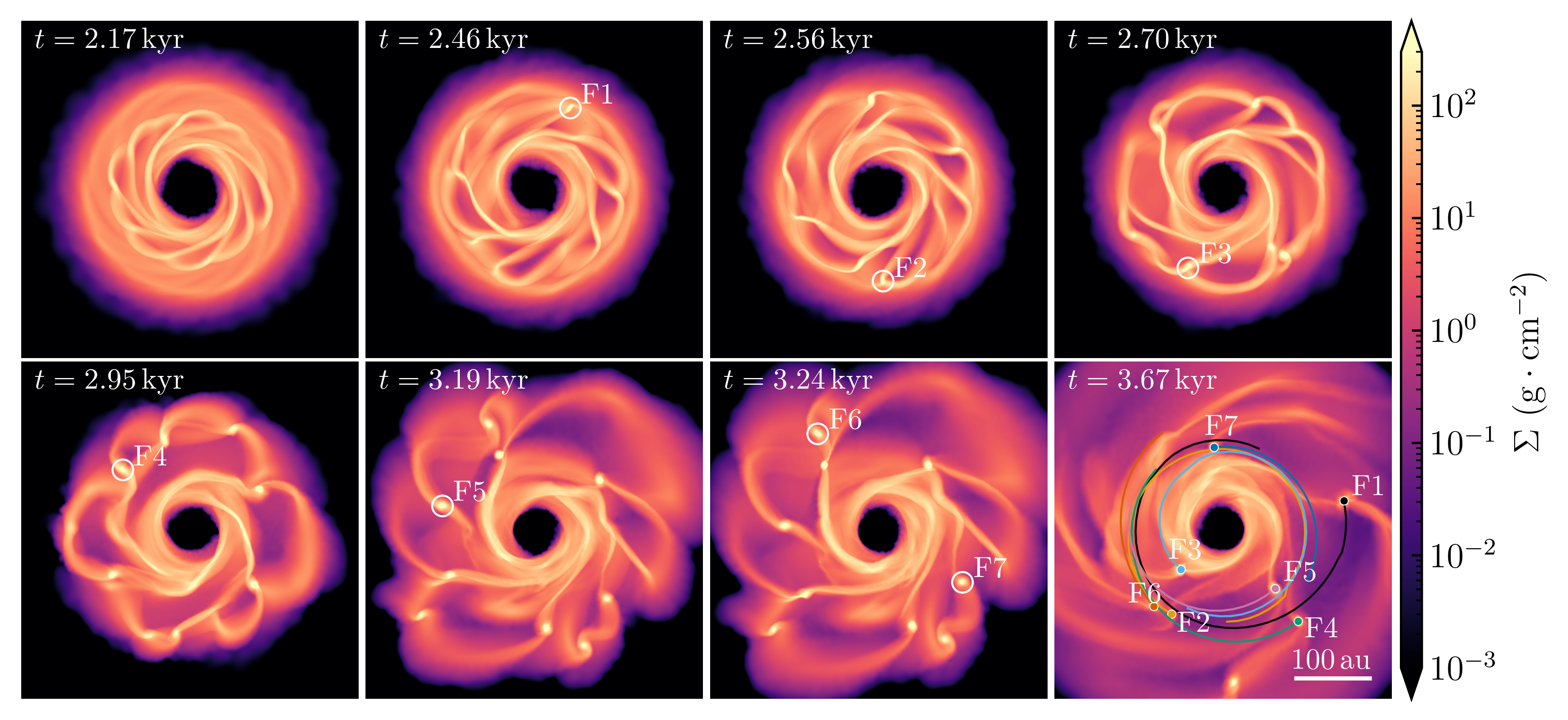}
\caption{Gas surface density before fragmentation ($t = 2.17$\,kyr, top
left), at the identification epoch of each fragment (each newly identified
fragment circled and labeled, and F6 and F7 sharing the $t = 3.24$\,kyr
panel),
and at the end of the run (bottom right), where the trajectories of the
seven fragments from birth to the end of the run are overlaid (colored
curves, labels at final positions). Note the transition from a smooth
ringed and spiral disk to the crowded, interaction-dominated late-time
state.
\label{fig:overview}}
\end{figure*}

\begin{figure*}[ht!]
\plotone{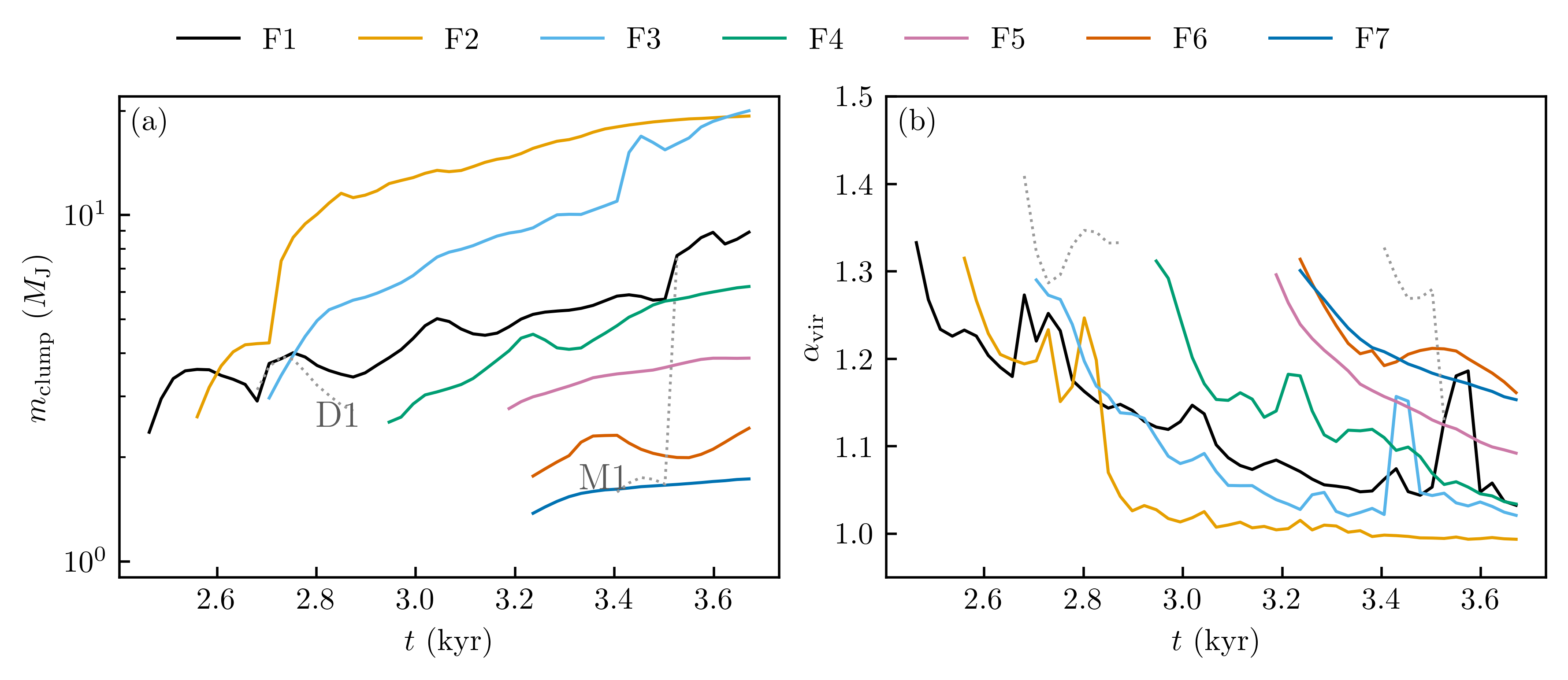}
\caption{Bound-clump census. (a) Mass evolution of all tracked clumps:
the seven fragments F1--F7 (colored solid curves) and the disrupted and
merging clumps D1 and M1 (gray dotted). M1's curve joins F1's at the merger.
(b) The virial parameter of Equation~(\ref{eq:alphavir}). Fragments descend
toward $\alpha_\mathrm{vir} = 1$ (minima $0.99$--$1.16$) while D1 and M1 turn
around at $\alpha_\mathrm{vir} \gtrsim 1.27$ before disruption or merger.
\label{fig:census}}
\end{figure*}

\begin{deluxetable*}{lccccccl}
\tablecaption{The bound-object census\label{tab:taxonomy}}
\tablehead{
\colhead{Object} & \colhead{$t_\mathrm{identify}$} &
\colhead{$m_\mathrm{frag}$} &
\colhead{$R_\mathrm{frag}$} & \colhead{$R_\mathrm{eff}$} &
\colhead{$\min\alpha_\mathrm{vir}$} & \colhead{$m(t_\mathrm{end})$} &
\colhead{Fate / notes} \\
\colhead{} & \colhead{(kyr)} & \colhead{($M_\mathrm{J}$)} &
\colhead{(au)} & \colhead{(au)} & \colhead{} & \colhead{($M_\mathrm{J}$)} &
\colhead{}
}
\startdata
F1 & $2.46$ & $2.36$  & $122$ & $4.4$ & $1.03$ & $8.9$ & survives, absorbs M1 at $3.53$\,kyr \\
F2 & $2.56$ & $2.62$ & $125$ & $4.0$ & $0.99$ & $19.3$ & survives, accretes debris \\
F3 & $2.70$ & $2.97$ & $117$ & $5.1$ & $1.02$ & $20.0$ & survives \\
F4 & $2.95$ & $2.53$ & $124$ & $4.5$ & $1.03$ & $6.2$ & survives \\
F5 & $3.19$ & $2.77$ & $130$ & $4.0$ & $1.09$ & $3.9$ & survives \\
F6 & $3.24$ & $1.77$ & $156$ & $4.1$ & $1.16$ & $2.4$ & survives, being scattered at $t_\mathrm{end}$ \\
F7 & $3.24$ & $1.38$ & $134$ & $3.2$ & $1.15$ & $1.7$ & survives \\
\hline
D1 & $2.68$ & $3.13$ & $126$ & $4.9$ & $1.28$ & \nodata & disrupted at $2.87$\,kyr \\
M1 & $3.41$ & $1.59$ & $177$ & $3.9$ & $1.27$ & \nodata & merges into F1 at $3.53$\,kyr \\
\enddata

\end{deluxetable*}

Figure~\ref{fig:overview} shows the disk fragmentation process and the evolution of representative clumps. The first bound clump, F1, emerges at
$t = 2.46\,\mathrm{kyr}$ at an orbital radius of $122$\,au, and eight
more bound clumps follow over the next kiloyear, the last of them M1 at
$3.41$\,kyr (Table~\ref{tab:taxonomy}). M1 survives for only $0.12$\,kyr before
merging into F1 at $3.53$\,kyr.
By the end of the run, the disk is crowded: seven fragments of
$1.7$--$20\,M_\mathrm{J}$ orbit between $79$ and $169$\,au, and their
gravitational interplay breaks the coherent spiral structure into segments. We note that several clumps (F1--F3) form when spirals collide with the insufficiently relaxed, smooth outer disk, rather than via the collapse of coherent spirals. This likely overestimates the number of fragments compared to simulations that start from a gravito-turbulent disk \citep{Deng_2017ApJ...847...43D}. However, the current setup provides a large sample of clumps to investigate.

Figure~\ref{fig:census} presents the census of the nine distinct bound clumps, and
Table~\ref{tab:taxonomy} lists their masses, identification epochs, and fates. Seven
of them survive to the end of the run (F1--F7, in order of formation). The two failures illustrate two distinct
channels. D1, born between F2 and F3, is disrupted after
$0.2\,\mathrm{kyr}$. M1 migrates inward until it merges with F1. The virial parameter separates the outcomes
cleanly (Figure~\ref{fig:census}b): every survivor descends toward
$\alpha_\mathrm{vir} = 1$, reaching minima of $0.99$--$1.16$, whereas D1 and M1
turn around at $\alpha_\mathrm{vir} \gtrsim 1.27$ before their
disruption or merger. This is the same threshold separating collapsing
fragments and dissolving clumps found in
\citetalias{Ni_2025ApJ...995...96N}.

\subsection{Initial fragment masses}\label{sec:initialmass}

\begin{figure}[ht!]
\plotone{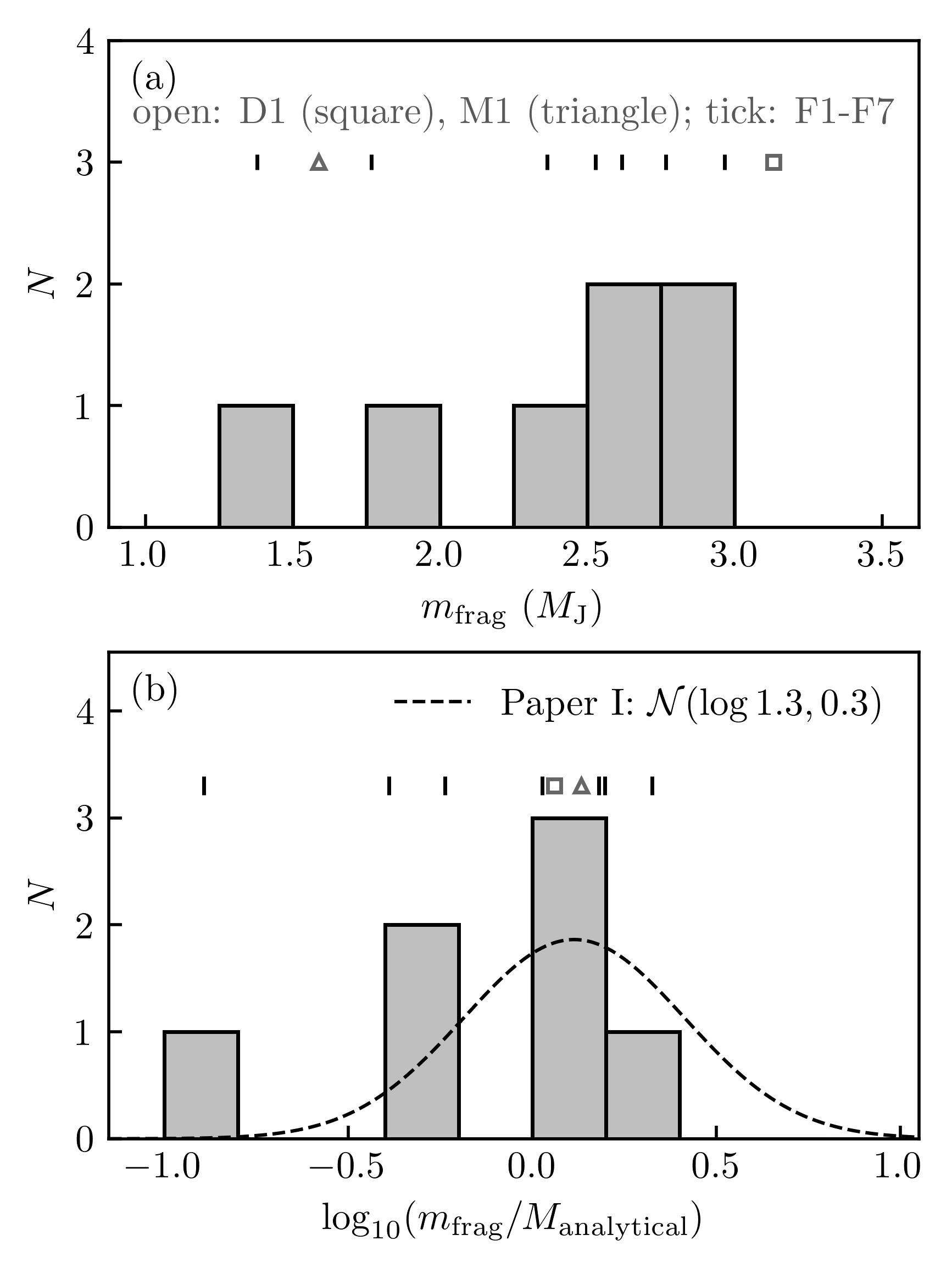}
\caption{Initial fragment masses. (a) Histogram of $m_\mathrm{frag}$ for
F1--F7, with ticks marking the individual fragments and open symbols
marking D1 (square) and M1 (triangle). (b) Initial masses normalized by
$M_\mathrm{analytical} = \Sigma \cdot \lambda_\mathrm{T}\cdot 2c_s/\Omega$
(Equation~(40) of \citetalias{Ni_2025ApJ...995...96N}, evaluated with the
isothermal sound speed and the slice averages defined in
Section~\ref{sec:initialmass}). The dashed curve is the log-normal found by
\citetalias{Ni_2025ApJ...995...96N}, scaled to $N = 7$. Ticks mark individual
fragments, and open symbols mark D1 and M1, which are not outliers in
initial mass.
\label{fig:initialmass}}
\end{figure}

Figure~\ref{fig:initialmass}a shows that the seven surviving fragments have initial masses of
$m_\mathrm{frag} = 1.4$--$3.0\,M_\mathrm{J}$. The failed objects are not separable by initial mass:
M1 ($1.6\,M_\mathrm{J}$) falls within the fragment range, and D1 ($3.1\,M_\mathrm{J}$) is only marginally above it.
Survival is therefore set by the post-formation dynamics, not by the initial collapse.

\citetalias{Ni_2025ApJ...995...96N} found that initial fragment masses follow the non-axisymmetric spiral-arm collapse estimate
$M_\mathrm{analytical} = \Sigma\,\lambda_\mathrm{T}\,(2c_s/\Omega)$, where
$\lambda_\mathrm{T} = 2c_s^2/(G\Sigma)$ is the most unstable (Toomre) wavelength. They reported that
$m_\mathrm{frag}/M_\mathrm{analytical}$ is approximately log-normally distributed with a central value of $1.3$ and a scatter of $0.3$\,dex.

We compute $M_\mathrm{analytical}$ using their prescription: $\Sigma$ and $c_s$ are azimuthally averaged over the radial slice
$R_\mathrm{frag} \pm R_\mathrm{eff}$ at $t_\mathrm{identify}$, excluding each fragment's own member particles, and we adopt the isothermal $c_s$.
Figure~\ref{fig:initialmass}b compares the measured ratios with this log-normal
distribution. The ratios have a median of $1.06$ and a standard deviation of
$0.43$\,dex. Seven fragments are too few for a meaningful statistical test, but we see
no inconsistency between the two.

\subsection{Accretion}
\label{sec:accretion}

\begin{figure*}[ht!]
\plotone{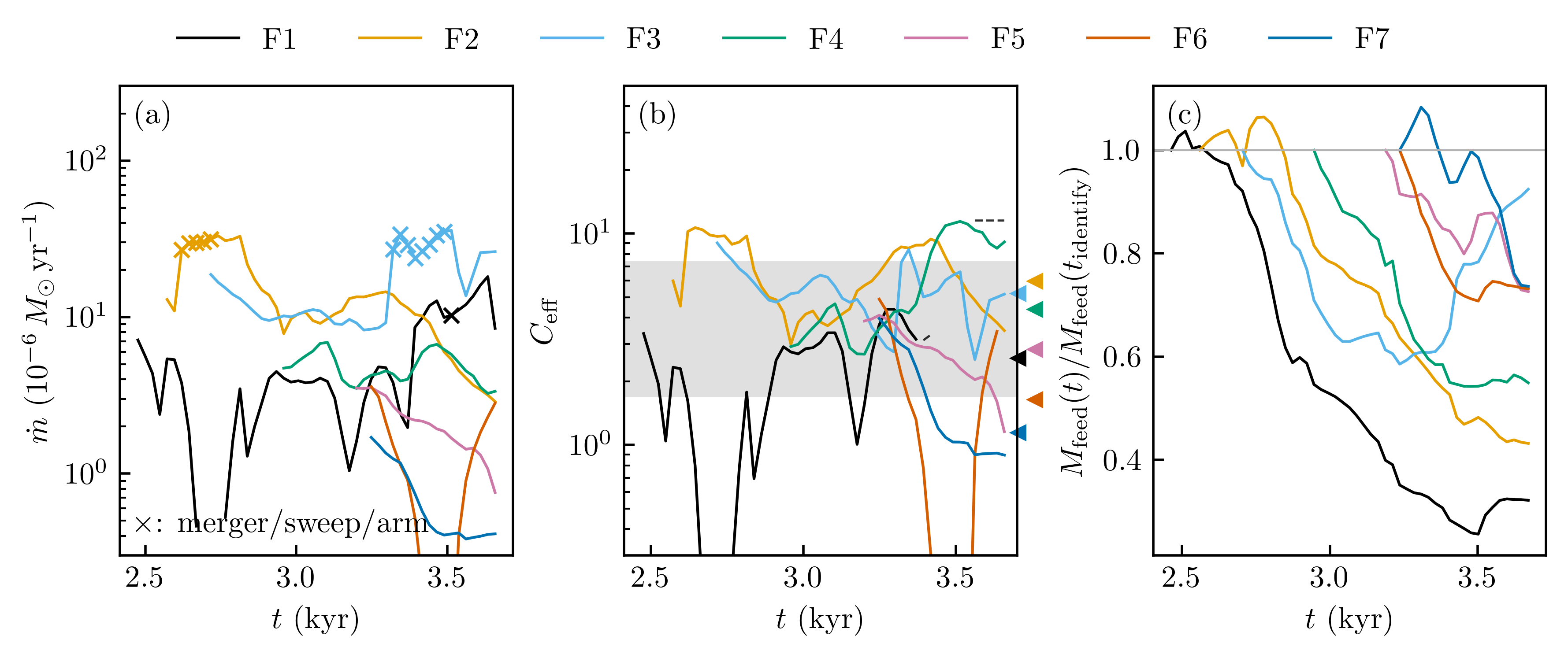}
\caption{Accretion history. (a) Net mass growth rate of each fragment, smoothed
over the orbital shear time (Section~\ref{sec:analysis}). No time intervals are
masked, so the episodes remain visible. Crosses indicate the three episodes
with identifiable mechanisms: F2's sweep-up of debris at $2.6$--$2.7$\,kyr, F3's
spiral-arm crossing at $3.3$--$3.5$\,kyr, and the merger of M1 into F1 at
$3.5$\,kyr. Brief intervals of negative smoothed growth (three for F1 near
$2.7$\,kyr and two for F6 during its scattering) are not shown on the logarithmic
axis. (b) Hill efficiency $C_\mathrm{eff}=\dot{m}/(\Sigma_\mathrm{loc}\Omega
R_\mathrm{H}^2)$. The shaded band shows the $16$th--$84$th percentile range over
all fragments and times, and left-pointing markers show per-fragment medians.
Intervals affected by the merger or by close encounters between clumps are
omitted. For F1 the curve is dashed from $3.4$\,kyr onward, once the averaging
window includes the merger, because the smoothed rate then reflects binding of
gas delivered by M1 rather than accretion from the disk. (c) Ambient mass
$M_\mathrm{feed}$ in each fragment's feeding annulus (Equation~(\ref{eq:mfeed})),
normalized to its value at the time the fragment is identified.
\label{fig:accretion}}
\end{figure*}

\begin{figure}[ht!]
\plotone{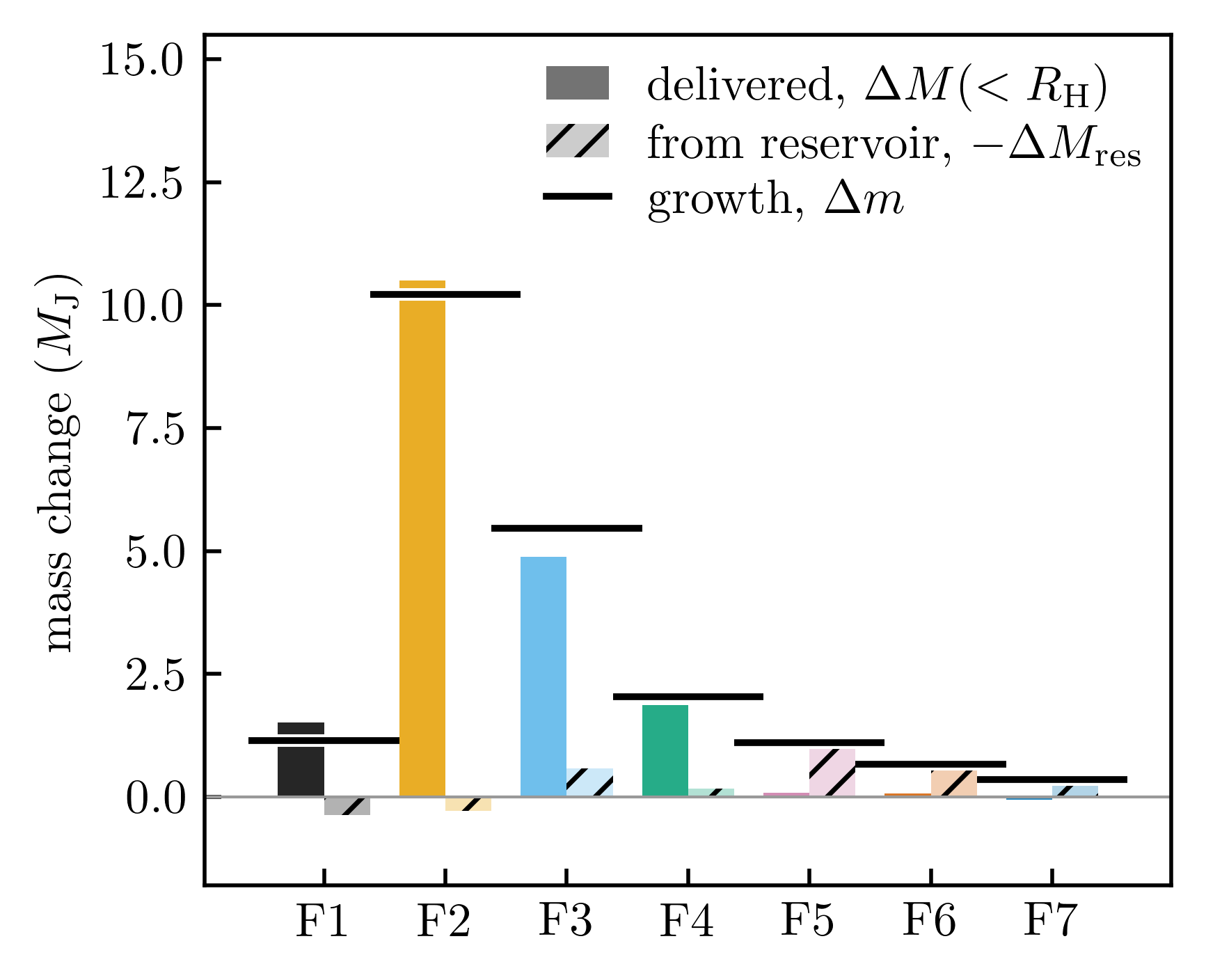}
\caption{Delivery and binding during the first $434.7$\,yr after each fragment is
identified (Section~\ref{sec:analysis}). Solid bars show the change in total gas
mass within the Hill sphere. Hatched bars show $-\,\Delta M_\mathrm{unbound}$,
i.e., the negative of the change in the unbound component: positive values mean
that the fragment depleted its unbound reservoir, while negative values (F1 and
F2) indicate that the unbound content increased. The black lines give the
measured change in bound mass, which (Equation~\ref{eq:twostep}) equals the sum
of the two bars as plotted. F1--F4 are primarily fed by
continued delivery, whereas F5--F7 grow mostly by binding gas already present
inside their Hill spheres. About a third of the growth of F2 in this interval comes
from sweeping up the debris of a transient overdensity at $2.6$--$2.7$\,kyr
(Section~\ref{sec:accretion}).
\label{fig:ledger}}
\end{figure}

We measure fragment growth by the time derivative of the bound mass, averaged over
the orbital shear time (Section~\ref{sec:analysis}). The averaging removes short-term
fluctuations, which arise because gas near the edge of a fragment repeatedly becomes
bound and unbound under the criterion of Equation~(\ref{eq:bound}).

Figure~\ref{fig:accretion}a shows the time-averaged growth rates. Two measured results set the natural rate scale. First, the thermal (Bondi) radius $R_\mathrm{B}$ exceeds the Hill radius $R_\mathrm{H}$
for every fragment at every epoch, with per-fragment median ratios
$R_\mathrm{B}/R_\mathrm{H}$ between $3.5$ and $10.5$. The stellar tide therefore limits
the region from which a fragment accretes to its Hill sphere, as found in local
simulations of embedded planets with $R_\mathrm{B} > R_\mathrm{H}$
\citep{Machida_2010MNRAS.405.1227M,Kuwahara_2019A&A...623A.179K,Bethune_2019MNRAS.488.2365B}.
Growth is thus Hill-limited. Second, $R_\mathrm{H}$ exceeds the local scale height $H$ by up to a factor of $\sim 2$, implying a two-dimensional feeding geometry and a natural rate scale $\Sigma_\mathrm{loc}\Omega R_\mathrm{H}^2$. With this choice, a single scaling describes the full population: the median efficiency is $C_\mathrm{eff} \approx 4$ for accretion rates spanning more than two decades, with per-fragment medians between $1$ and $6$ (Figure~\ref{fig:accretion}b). Growth thus proceeds at a few times the nominal Hill rate, sustained by the dense gas in which the fragments remain embedded. This should be interpreted as an effective scaling for the time-averaged increase of the bound mass, not as an instantaneous mass flux across the Hill surface.

We describe accretion as a two-step process. Gas is first \emph{delivered} from the surrounding disk into the Hill sphere, and only then becomes \emph{bound} to the fragment. The Hill sphere therefore serves as an intermediate reservoir: growth is delivery-limited if it is not replenished as quickly as it is depleted, and binding-limited otherwise. Defining the reservoir mass as $M_\mathrm{res} \equiv M(<R_\mathrm{H}) - m$, the bound-mass change over any interval obeys
\begin{equation}\label{eq:twostep}
    \Delta m = \Delta M(<R_\mathrm{H}) - \Delta M_\mathrm{res}.
\end{equation}
Thus the net bound-mass growth decomposes into (i) the change in the total
Hill-sphere inventory (delivery) and (ii) the change in the unbound reservoir
(binding). Figure~\ref{fig:ledger} shows both terms in
Equation~(\ref{eq:twostep}) for all seven fragments. In broad terms, the early
clumps (F1--F4) efficiently bind newly delivered gas, whereas later fragments
grow mainly by consuming the gas already present within their Hill spheres.

Specifically, over the first $434.7$\,yr after identification, the Hill spheres of the early
fragments (F1--F4) gain nearly as much mass as the fragments themselves
($1.5$--$10.5\,M_\mathrm{J}$ versus $1.1$--$10.2\,M_\mathrm{J}$). The disk therefore
replenishes the reservoir about as quickly as binding depletes it, so delivery
is not rate limiting. In contrast, the Hill spheres of the late fragments
increase by $<0.1\,M_\mathrm{J}$ over the same interval while their bound masses
grow by $0.35$--$1.1\,M_\mathrm{J}$. Their growth is drawn from gas already
present inside $R_\mathrm{H}$, whose inventory declines by a comparable amount:
these fragments consume a finite reservoir rather than being actively fed. For
F5--F7, this interval is essentially their full lifetime to the end of the run.

This delivery contrast reflects environment and formation history. The late
fragments condensed from spiral arms in a disk that had already been partially
drained by the earlier generation. At fixed age (Section~\ref{sec:analysis}),
their surroundings are markedly more rarefied: the median ambient density in a shell from $1.2$ to $2.5\,R_\mathrm{H}$ is lower by a factor of $\sim 5$, the local
surface density is lower by a factor of $\sim 2$, and the total inflow into the
Hill sphere is reduced by more than an order of magnitude relative to the early
fragments. This shortfall is self-reinforcing, because weaker delivery leaves
less mass and hence a smaller Hill cross section, further reducing subsequent
access. The late fragments therefore remain poorly supplied to the end of the run. Our disk fragments before it reaches a gravito-turbulent state (Section~\ref{sec:census}), so the early fragments form and grow in dense, rapidly evolving spiral arms. It is possible that a disk that fragments from a gravito-turbulent state, where the spiral arms are quasi-steady, supplies lower accretion rates to its fragments.

The mass delivery and long-term mass growth are limited by the local mass supply. For F1--F4 we trace each delivered particle back five snapshots, i.e., $121$\,yr, prior to its first
crossing of the Hill surface, and measure its radial offset from the fragment's
orbit (in units of $R_\mathrm{H}$) at that earlier time. Half of the delivered gas
originates within $1.3$--$1.6\,R_\mathrm{H}$ of the orbit, and $95$\% within
$2.0$--$2.4\,R_\mathrm{H}$. This measured feeding width motivates the $a\pm2R_\mathrm{H}$
feeding annulus adopted in Section~\ref{sec:analysis}.

These annuli are being exhausted. Each fragment's annulus drains, ending at
$32$--$92$\% of its mass at identification (Figure~\ref{fig:accretion}c). The
fragments themselves account for much of this depletion: by the end of the run
the bound objects contain $62\,M_\mathrm{J}$, or $31$\% of the simulation's total
gas mass, and each newly formed fragment competes for the same local supply.
Meanwhile, global flows driven by the coupled disk--fragment dynamics
redistribute the remaining gas, temporarily refilling some annuli while draining
others (e.g., F7 briefly exceeds its initial value). Growth is therefore
supply-limited on the disk-evolution timescale, $10^{-1}$--$10^{0}$\,Myr, and a
fragment's long-term outcome is set less by its instantaneous accretion rate
than by what the disk can continue to deliver. Section~\ref{sec:finalmass}
quantifies this prediction.

Discrete episodes punctuate the growth of the early fragments. F2 accretes
$+3.7\,M_\mathrm{J}$ of dense debris between $t=2.61$ and $2.73$\,kyr, sourced from a
transient overdensity that dispersed shortly after fragmentation began. F3
crosses a spiral arm between $t=3.31$ and $3.50$\,kyr and gains
$+5.4\,M_\mathrm{J}$. The merger with M1 adds $+1.9\,M_\mathrm{J}$ to F1 at
$t=3.5$\,kyr. Because the growth rate is averaged over a $\sim270$\,yr window, the
merger influences F1's inferred efficiency for roughly half a window on either
side; consequently, the dashed segment in Figure~\ref{fig:accretion}b begins near
$3.4$\,kyr rather than exactly at the merger time.

\subsection{Migration}
\label{sec:migration}

\begin{figure*}[ht!]
\plotone{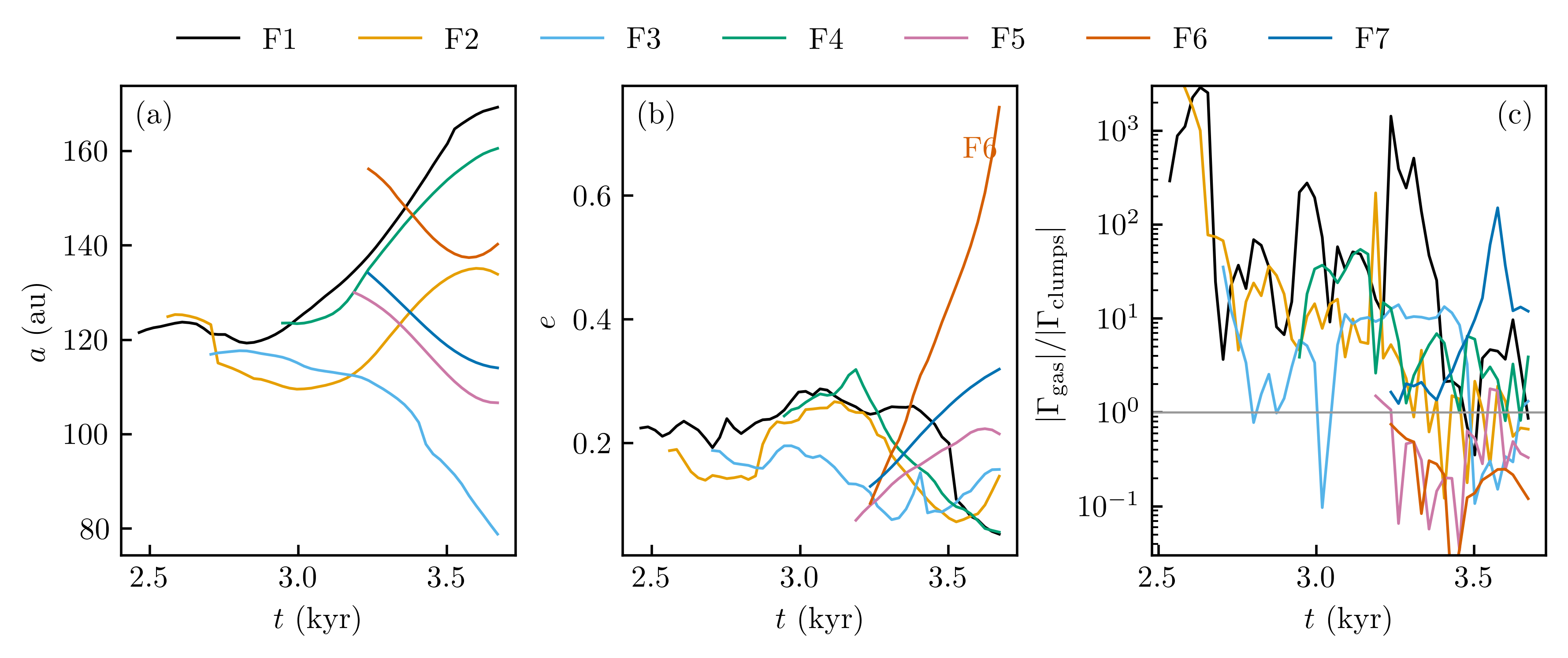}
\caption{Migration. (a) Orbital radius of each fragment. (b) Orbital
eccentricity, with the labeled curve showing F6, which is being scattered
when the run ends. (c) Ratio of the gravitational torque from ambient gas to that
from the other clumps (Section~\ref{sec:analysis}), on the bound members
of each fragment. The horizontal line marks equality.
\label{fig:migration}}
\end{figure*}

Figure~\ref{fig:migration}a shows the orbital histories. Migration is rapid and
bidirectional: three fragments migrate outward on average (F1, F2, and F4),
while four migrate inward (F3, F5, F6, and F7), with mean rates up to
$0.05\,\mathrm{au\,yr^{-1}}$.
For the three most recently formed fragments, tracked for only
$\sim\!0.25$--$0.4$ orbits, the mean inward drift is comparable to the epicyclic
radial excursions of their eccentric trajectories; their long-term migration
sense is therefore less certain than for the longer-lived fragments.
Relative to the linear Type-I timescale $t_\mathrm{I}$ (Section~\ref{sec:analysis}),
only F3 migrates as expected, inward with $|a/\dot a|/t_\mathrm{I}=0.9$.
For the other fragments, the median ratios span $0.1$--$1.6$, implying migration
that is up to an order of magnitude faster than the linear rate, or even
outward, contrary to linear-theory expectations. Because the fragments are
super-thermal and the disk is neither laminar nor smooth, we use this
comparison only as a reference scale, not as a test of the theory.

The angular-momentum budget clarifies what drives the drift. For each fragment,
the net change in total angular momentum equals the angular momentum accreted by
particles that join the bound set, minus that removed by particles that leave it, plus the
external gravitational torque from the ambient gas and from the other clumps.
Here, we focus on the latter two terms to identify the dynamical driver of the
orbital evolution. Using Equation~(\ref{eq:torque}), Figure~\ref{fig:migration}c
decomposes the external torque into contributions from gas and from neighboring
clumps. For five fragments (F1--F4 and F7), the gas torque dominates; for F5 and
F6, interactions with other clumps dominate, so their migration is governed by
few-body dynamics rather than disk torques.
Fragment F6 is the clearest example: its nearest neighbor is F2, which first
recedes and then steadily approaches, while F6's eccentricity grows from its
identification at $t=3.24$\,kyr to $e=0.74$ at the end of the run. The scattering is still underway when the simulation ends.

\subsection{Spin}\label{sec:spin}

\begin{figure*}[ht!]
\plotone{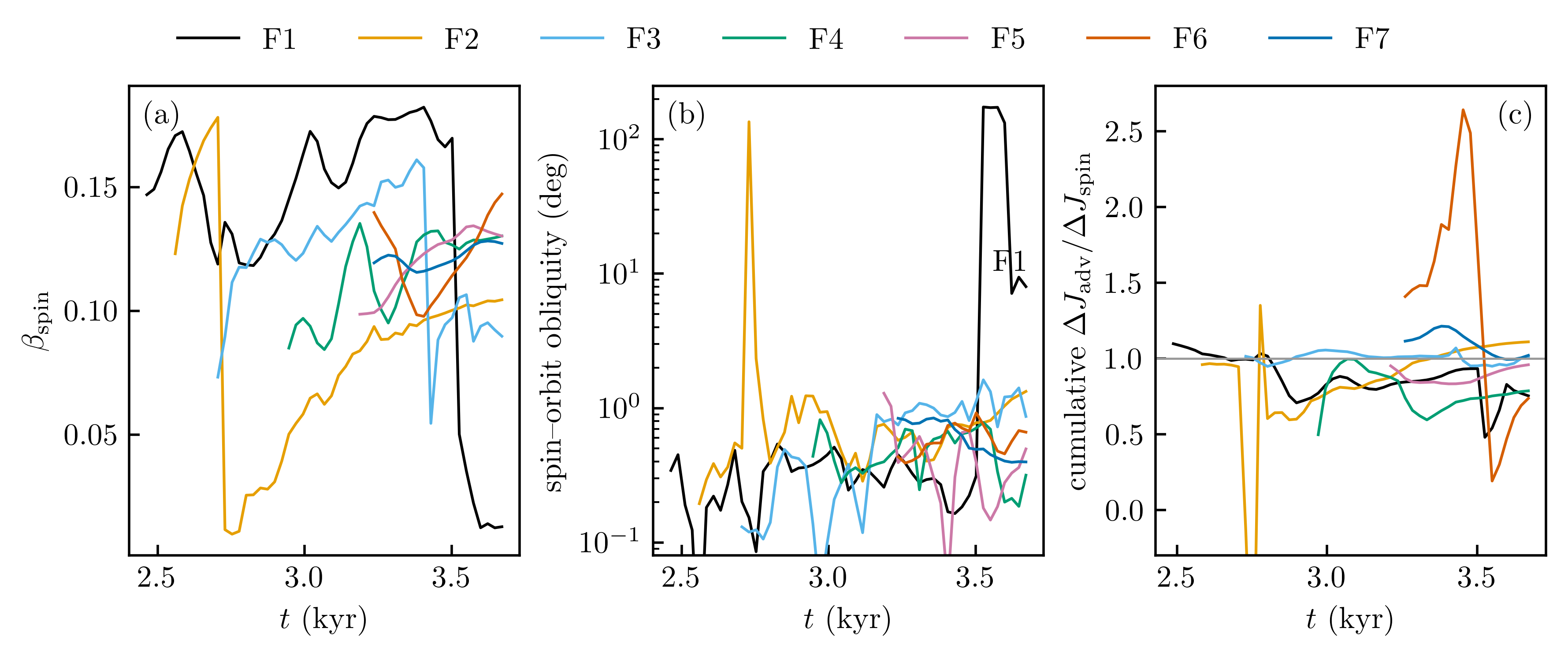}
\caption{Spin diagnostics for the fragments. (a) Rotational support,
$\beta_\mathrm{spin}$ (Section~\ref{sec:analysis}). (b) Spin--orbit
obliquity on a logarithmic scale; the post-merger fragment F1 is labeled.
(c) Running total of the net spin angular momentum advected by joining
particles minus that removed by leaving particles, shown as a fraction of each
fragment's spin change accumulated to the same time
(Section~\ref{sec:analysis}). The horizontal line marks unity.
\label{fig:spin}}
\end{figure*}

Figure~\ref{fig:spin} summarizes the fragments' rotational state. The six fragments
untouched by the merger spin up overall, with brief reversals and
obliquity excursions during accretion episodes (most clearly for F2 and
F3 in Figure~\ref{fig:spin}a and b).

By the end of the run, their rotational support spans
$\beta_\mathrm{spin}=0.09$--$0.15$. In general, these are
pressure-supported bodies that rotation flattens only mildly.

Their rotation is differential and largely ordered. The mean azimuthal
velocity about the spin axis exceeds its fluctuations by factors of a
few. The angular velocity is nearly uniform in the innermost shells and declines
increasingly steeply farther out. Between the radii enclosing one quarter and one half
of the mass, its power-law index $q_\Omega$ lies between $-1.09$ and $-0.12$ for the
six unmerged fragments. For comparison, $q_\Omega = 0$ for solid-body rotation and
$q_\Omega = -3/2$ for Keplerian rotation.
The spins are also well aligned with the orbital angular momentum, with
spin--orbit obliquities mostly below $1^{\circ}$ (Figure~\ref{fig:spin}b).

Accretion supplies most of this spin. Integrated over each fragment's
life, the net spin angular momentum advected by joining particles (minus that
removed by leaving particles) accounts for most of the spin change.
Figure~\ref{fig:spin}c shows this running total, which fluctuates more at
early times because its denominator---the accumulated spin change---is
still small.

F1 is the exception. The merger leaves it slowly rotating
($\beta_\mathrm{spin}=0.013$), with much less ordered rotation, and tilted. Its spin axis is inclined by $8^{\circ}$ to
the orbital axis, giving the only final obliquity above $1.5^{\circ}$ in
the sample. Thus a single merger---here with mass ratio $0.29$ (M1/F1)
immediately prior to coalescence---can reset the spin state that accretion
had built.

\subsection{Interior structure}
\label{sec:interior}

\begin{figure*}[ht!]
\plotone{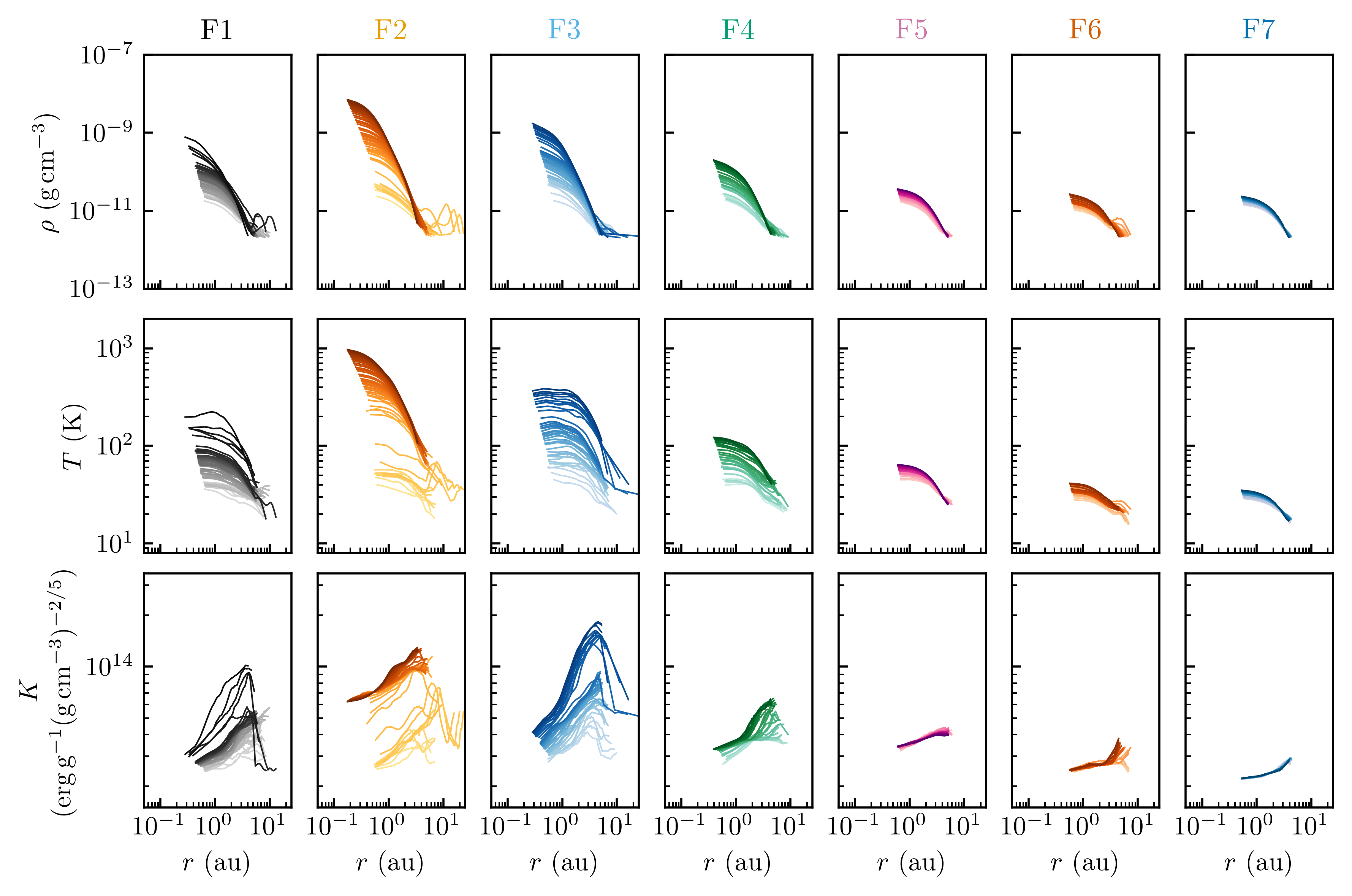}
\caption{Interior profiles of all seven fragments at every analyzed
epoch: density (top row), temperature (middle), and adiabat label
$K = (\gamma-1)\,u\,\rho^{1-\gamma}$ (bottom, in units of
$\mathrm{erg\,g^{-1}(g\,cm^{-3})^{-2/5}}$), as shell medians about each
fragment's density peak (Section~\ref{sec:analysis}). Within each column
the line color runs from light (identification) to dark (end of the run)
in a per-fragment color sequence. In the bottom row the final,
monotonic profiles of F1 and F3 are partly overlain by the immediately
preceding post-episode epochs, drawn at nearly the same dark color.
\label{fig:interiorevo}}
\end{figure*}

\begin{figure}[ht!]
\plotone{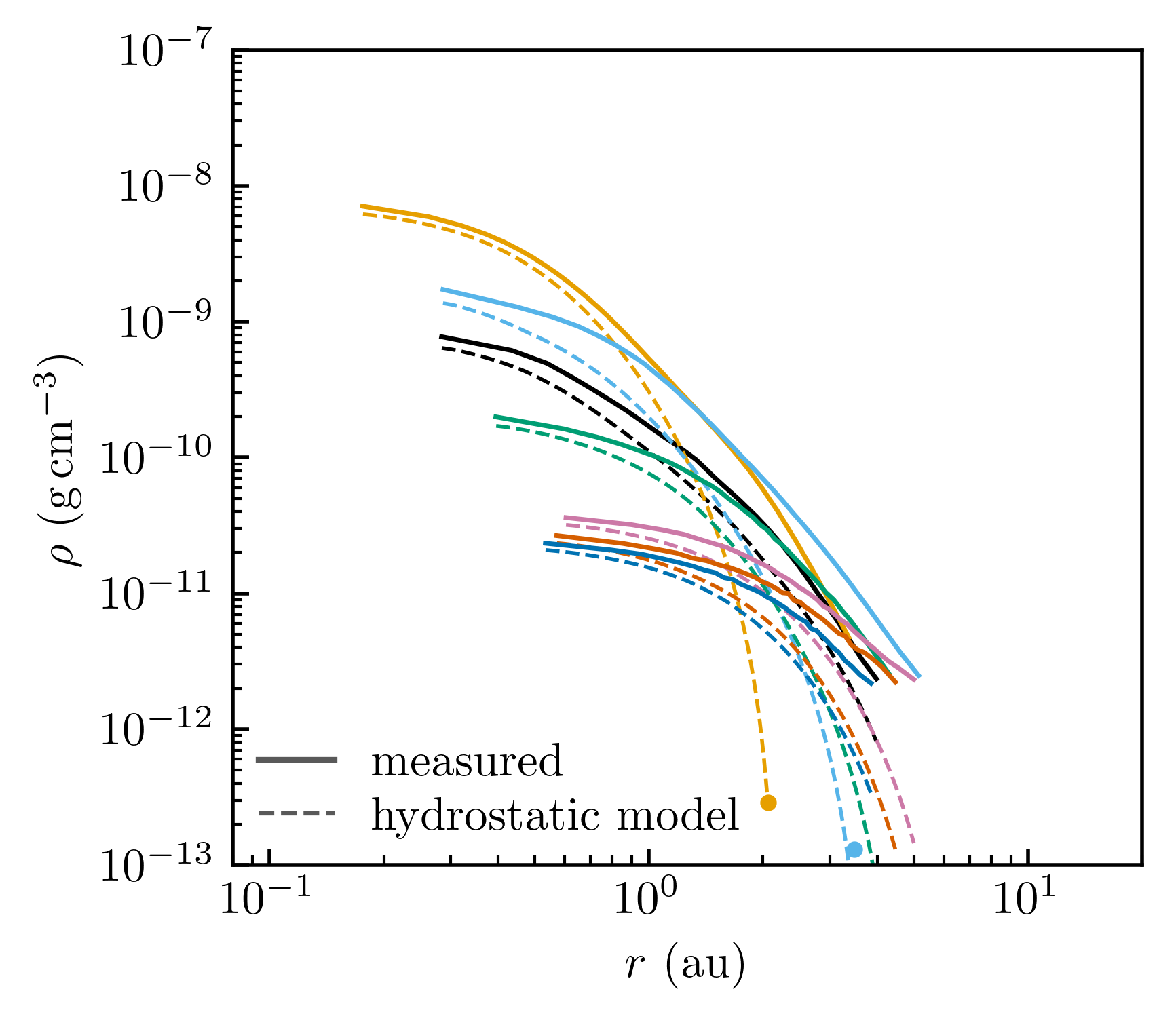}
\caption{Measured density profiles of all seven fragments at the
end of the run (solid), compared with hydrostatic models integrated from
each fragment's measured $K(m_r)$ and central density (dashed). The dashed curves are clipped to each fragment's
measured radial range, and a dot marks the radius where a model reaches
zero pressure, drawn only for F2 and F3, whose surfaces fall inside this
range. Colors are as in Figure~\ref{fig:census}.
\label{fig:strat}}
\end{figure}

Figure~\ref{fig:interiorevo} presents the density, temperature, and entropy
profiles of each fragment at all analyzed epochs. Despite forming at
different times, all seven follow the same structural and evolutionary
pattern, with a centrally condensed core a few astronomical units across
embedded in an extended envelope. Over each fragment's lifetime the
central density increases by up to $\sim\!2.5$ orders of magnitude, while
the central temperature rises from $28$--$44$\,K at identification to
$35$--$990$\,K by the end of the run. The heating comes mainly from $P\,\mathrm{d}V$ work done as the accreted mass
compresses the interior. The fragments that accrete the most heat the most, with F2
reaching $\sim\!990$\,K, whereas the late-forming fragments warm only slightly. The adiabat label $K$ generally
increases outward through the inner $\sim$half of the bound mass at every
epoch, implying that fragments are born entropy-stratified and become
increasingly resistant to convection as they grow, with most of the
interior Schwarzschild-stable by the end. Finally, thermal energy provides
the dominant support throughout. The ratio $U/|E_\mathrm{P}|$ decreases from
$0.72$--$0.87$ at identification to $0.63$--$0.74$ at the end of the run.

If a fragment's interior is a hydrostatic structure supported by the pressure of gas at
its measured entropy, then integrating the equation of hydrostatic
equilibrium in the interior mass coordinate $m_r$, which is the bound
mass enclosed within radius $r$,
\begin{equation}\label{eq:hse}
    \frac{\mathrm{d}P}{\mathrm{d}m_r} = -\frac{G m_r}{4\pi r^{4}},
    \quad
    \frac{\mathrm{d}r}{\mathrm{d}m_r} = \frac{1}{4\pi r^{2}\rho},
    \quad \rho = \left[\frac{P}{K(m_r)}\right]^{1/\gamma},
\end{equation}
with each fragment's own measured $K(m_r)$ and central density, must
reproduce its measured density profile. Figure~\ref{fig:strat} shows that the models generally do. The innermost mass is closest to thermal hydrostatic balance, since ordered rotation, measured as the shell average of $\langle v_\phi\rangle^{2} r/(G m_r)$ over the inner half of the mass, supplies only about a tenth to a quarter of the support there, while in the envelope, rotation and the unsettled motion of
newly arrived gas contribute support that a purely thermal model
omits. Taken together, the approximately
matched inner region and the
underestimated outer part suggest a centrally condensed,
entropy-stratified interior in thermal hydrostatic balance, surrounded by an
accreting, rotating envelope that is not.

The hierarchy of the three timescales defined in Section~\ref{sec:analysis} determines the
physics here. Across the seven fragments, $t_\mathrm{ff} \sim
10^{0}$--$10^{1}$\,yr, $t_\mathrm{acc} \sim 10^{3}$\,yr, and
$t_\mathrm{KH} \sim$ a few $\times\,10^{4}$\,yr. Because
$t_\mathrm{ff} \ll t_\mathrm{acc}$, a fragment approaches hydrostatic
equilibrium quickly while it grows. On the other hand, because $t_\mathrm{acc} \ll t_\mathrm{KH}$, it
grows much faster than it can cool, so its structure is mostly shaped by
accretion, and the compression that accretion
drives is nearly adiabatic: a parcel of gas largely keeps the entropy it
had when it was bound, since convection is suppressed by the
stratification itself and the opaque interior lets out only a few percent
of the accretion power. The entropy profiles in the bottom row of
Figure~\ref{fig:interiorevo} can then be read as a record. In
space, the outward rise of $K$ traces a stack in which the deepest layers are the earliest, lowest-entropy
arrivals and each new layer settles on top with slightly higher entropy,
so the amplitude of the stratification implies how much the fragment
has accreted. The outermost parts of the fragments develop transient
entropy-decreasing layers, where gas assembly is entangled with the
complicated dynamics of the surrounding disk. The deepest such
layers appear during and just after a growth episode, namely F1's
merger, F2's debris sweep and F3's arm crossing, with shallower
inversions in F3's first epochs while its initial envelope settles. In time, the interior curves of the starved
late fragments lie on top of one another from epoch to epoch, frozen
because little is being added, while the profiles of the strongly fed
early fragments extend outward and rise at their outer ends as accretion
stacks new layers, severalfold for F2 and F3. Only the interiors of the
strongest accretors drift slowly upward, the residue of irreversible
heating during their most rapid growth.

The internal structure has two implications for the evolution of these objects. The
first concerns their thermal state, and it bears on the distinction
between the hot-start and cold-start initial conditions used in models
of giant-planet evolution, which differ in how much of the entropy of
the accreted gas a young object retains
\citep[e.g.,][]{Burrows_1997ApJ...491..856B, Marley_2007ApJ...655..541M, Spiegel_2012ApJ...745..174S, Marleau_2014MNRAS.437.1378M}. That distinction is observable,
because a hot start leaves an object markedly more luminous at a given
mass and age. In the final profiles the entropy rises outward, so a fragment's central value is the lowest it holds. In our simulation, the specific entropies at the center at the end of the
run are $s = $ $12.0$, $12.9$, $12.3$, $12.2$, $12.2$, $11.9$ and
$11.8$\,$k_\mathrm{B}$ per baryon for F1 to F7. All lie well above the
$9.2$ to $10.5$\,$k_\mathrm{B}$ per baryon that those models require of the
directly imaged planets \citep{Marleau_2014MNRAS.437.1378M}: values that imply
hot-start initial conditions should these objects eventually become gas
giants. We will discuss their fates in Section~\ref{sec:finalmass}.

The second concerns survival. The surviving interiors are robust
against the stellar tide. A clump that outgrows its Hill sphere is disrupted by the stellar tide \citep{Zhu_2012ApJ...746..110Z,Kratter_2016ARA&A..54..271K}. By the end of the run the mean density inside each fragment's half-mass radius exceeds $9M_\star/(4\pi a^{3})$, the mean density of a body that fills its Hill sphere, by a factor between $40$ and $1.0\times10^{4}$. The cores built by this quasi-adiabatic compression are
therefore effectively immune to destruction by the stellar tide at their
present orbits. Survival is decided by the dynamical environment, as
Section~\ref{sec:census} showed.

\subsection{Final mass estimates}\label{sec:finalmass}

\begin{figure}[ht!]
\plotone{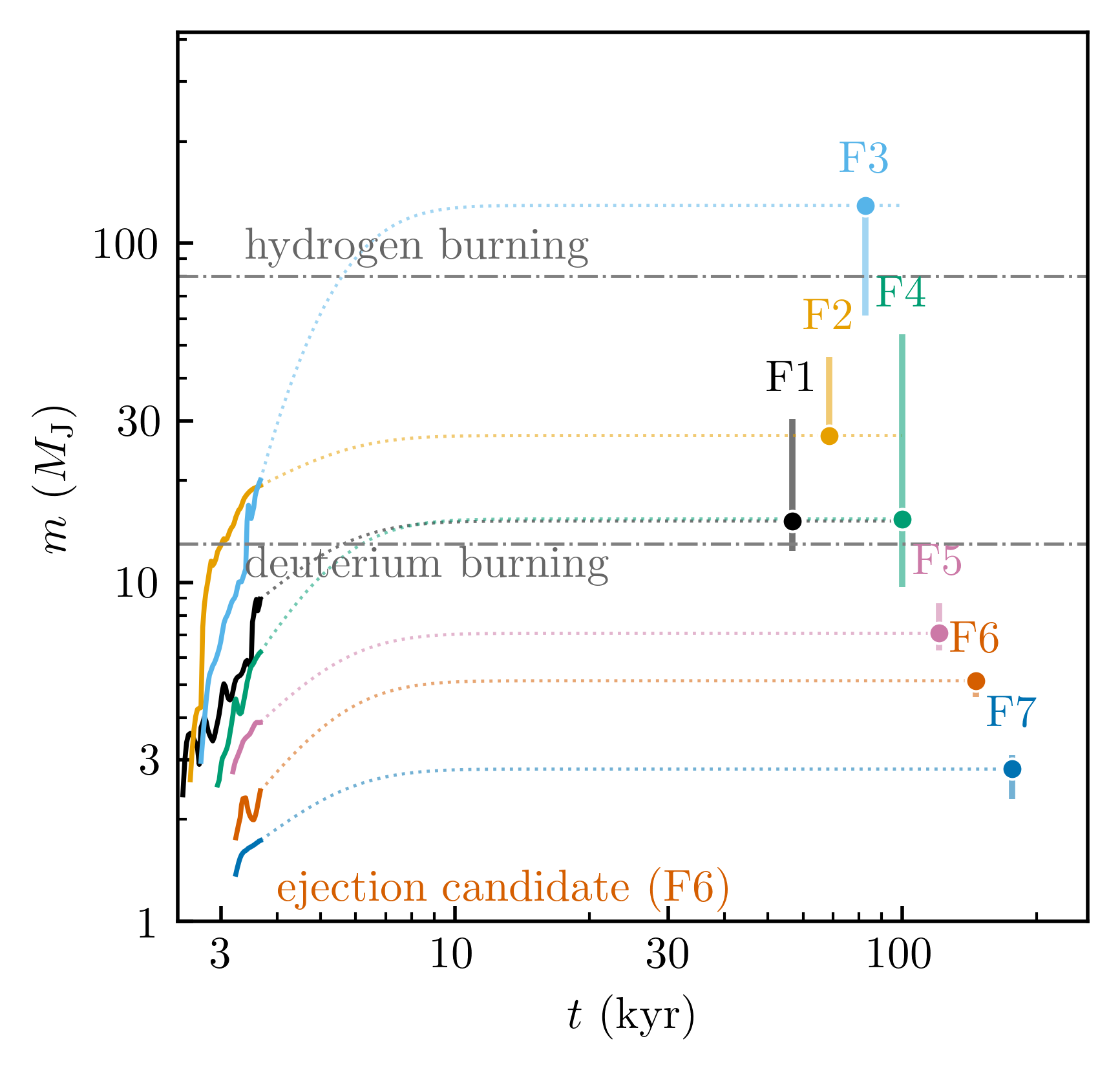}
\caption{Final masses. Solid curves are the measured mass
histories, on a logarithmic time axis, and dotted curves continue each
fragment along its computed mass evolution, which saturates once the
shared reservoir is consumed. Filled
circles mark the conditional masses at $0.1$\,Myr from the calculation of
Section~\ref{sec:finalmass}. The thin bars span the adopted value
together with the variants in which
the lifetime-median surroundings, the lifetime-median efficiencies, or
F1's literal
end-of-run efficiency separately replace the adopted \mbox{end-state} inputs. These are
assumption sensitivities, not confidence intervals, and the markers are
offset horizontally for legibility. The labeled annotation marks F6,
whose ongoing scattering may end in ejection. Dash-dotted lines mark
the deuterium-burning threshold and the hydrogen-burning boundary.
\label{fig:finalmass}}
\end{figure}

The run does not directly give a unique final mass for each fragment, because the simulated span is short compared to the disk lifetime of $10^{-1}$--$10^{0}$\,Myr. F2 and F3 exceed the deuterium-burning threshold
($13\,M_\mathrm{J}$), and competitive accretion determines which fragments can remain in the planetary mass regime. The simulation holds
$202\,M_\mathrm{J}$ of gas, of which $62\,M_\mathrm{J}$ is already in
the seven fragments and $140\,M_\mathrm{J}$ is ambient
($88\,M_\mathrm{J}$ of it between $60$ and $200$\,au), with only
$8.6\,M_\mathrm{J}$ of ambient gas currently inside the seven Hill
spheres (each particle assigned to the fragment for which its distance in
units of that fragment's Hill radius is smallest). We therefore report \emph{conditional}
survivor masses. We integrate the growth law forward for every
fragment at once, drawing on a single shared reservoir,
\begin{equation}\label{eq:forward}
    \frac{\mathrm{d}m_i(t)}{\mathrm{d}t} =
    C_i\,\Sigma_i(t)\,\Omega_i(t) \,R_{\mathrm{H},i}^{2}(t),
\end{equation}
\begin{equation}\label{eq:pool}
    \frac{\mathrm{d}M_\mathrm{pool}(t)}{\mathrm{d}t} =
    -\sum_i \frac{\mathrm{d}m_i(t)}{\mathrm{d}t},
\end{equation}
where $C_i$ is an assumed constant efficiency of order unity, the
scale set by the measured values of Section~\ref{sec:accretion}, and
$M_\mathrm{pool}$ is the accessible reservoir, meaning the part of the
ambient gas that the disk can deliver to the fragments. The Hill radius
follows the growing mass while the orbital frequency follows the orbit,
\begin{equation}\label{eq:forwardgeom}
    R_{\mathrm{H},i}(t) = a_i(t)
        \left[\frac{m_i(t)}{3M_\star}\right]^{1/3},
    \qquad
    \Omega_i(t) = \left[\frac{GM_\star}{a_i(t)^{3}}\right]^{1/2},
\end{equation}
and the surface density available to each fragment declines as the
disk is drained,
\begin{equation}\label{eq:sigdecay}
    \Sigma_i(t) = \Sigma_i(0)\,M_\mathrm{pool}(t)/M_\mathrm{pool}(0).
\end{equation}
This closure is ad hoc: it ties the surface density around each
fragment to the overall disk mass evolution by assuming uniform depletion.
In addition, the disk gas may not all get accreted by the fragments, so we set
$M_\mathrm{pool}(0) = f_\mathrm{acc} M_\mathrm{amb}$, with
$M_\mathrm{amb} = 140\,M_\mathrm{J}$ the measured ambient mass and
$f_\mathrm{acc}$ the accessible fraction.

Three inputs remain undetermined: the efficiencies $C_i$, the orbits $a_i(t)$, and the accessible fraction $f_\mathrm{acc}$. For simplicity, we hold each orbit fixed at its end-of-run radius and set $f_\mathrm{acc}=1$, i.e., we assume the full current gas budget is available to the fragments and that no new gas arrives, e.g. from later infall.

We set each $C_i$ to the median efficiency measured over the fragment's last five clean intervals: $C_i = 4.1$, $4.1$, $4.8$, $9.1$, $1.9$, $1.8$, and $0.9$ for F1--F7. For F1 we use the last intervals unaffected by the merger, because its literal end-of-run value ($11.5$) reflects the binding of M1's debris rather than accretion from the disk (Section~\ref{sec:accretion}). For the starved F5--F7, these efficiencies mainly capture the binding of gas already resident within their Hill spheres rather than fresh delivery; if anything, the model therefore overstates their draw on the shared pool.

Each integration starts from the measured end state of every fragment---its bound mass $m_i(0)$, orbital radius $a_i(0)$, and a starting surface density $\Sigma_i(0)$ set to the median local $\Sigma_\mathrm{loc}$ over the last five snapshots---and advances to $0.1$\,Myr using a fourth-order Runge--Kutta scheme with $2500$ steps. At this resolution, the final masses change by $<0.01$\% when the number of steps is doubled.

Figure~\ref{fig:finalmass} shows the outcome. Because the supply declines only as the shared pool is depleted, accretion continues until the pool is exhausted; in this run it is exhausted rapidly, with $99$\% consumed within $7.2$\,kyr. Each fragment's share tracks its initial rate $C_i\,\Sigma_i(0)\,\Omega_i\,R_{\mathrm{H},i}^{2}(0)$. F3, which combines the densest end-of-run surroundings with the largest bound mass, takes $78$\% of the ambient budget and reaches $130\,M_\mathrm{J}$, exceeding the hydrogen-burning boundary ($\approx 80\,M_\mathrm{J}$) and thus becoming a low-mass stellar companion under these assumptions. F2 settles at $27\,M_\mathrm{J}$ and remains a brown dwarf. F1 and F4 end just above the deuterium threshold at $15\,M_\mathrm{J}$ each, while the late fragments F5--F7 remain gas giants at $7.1$, $5.1$, and $2.8\,M_\mathrm{J}$.

These masses are conditional on the stated assumptions, and their robustness varies. Separately replacing either the end-state surroundings or the efficiencies with lifetime medians shifts individual masses by factors of a few (e.g., F3 between $61$ and $133\,M_\mathrm{J}$; F4 between $10$ and $54\,M_\mathrm{J}$). In every variant, F2 remains a brown dwarf, F5--F7 remain giants, and F3 is the most massive object. However, in the lifetime-median-surroundings variant F4's gain slightly exceeds F3's, and the deuterium classification of F1 and F4 is not robust because their variants straddle the threshold. The two caveats that follow therefore bear mainly on F3.

The first caveat is gap opening. For the dominant consumer, its effect is one-sided: a gap around F3 would reduce F3's growth, whether the withheld gas is redistributed through the shared pool or never delivered at all. All fragments already satisfy the thermal gap-opening criterion, with end-of-run Hill radii exceeding the local scale height by factors of $1.0$--$2.1$; in a low-viscosity, laminar disk, objects this massive would clear their co-orbital gas and throttle their own supply \citep{Crida_2006Icar..181..587C}.

Two measured properties of this disk work against clean gaps. First, the torques arise from spiral structure and neighboring clumps rather than from a smooth viscous background (Figure~\ref{fig:migration}c). Second, the orbits are eccentric: the end-of-run apocenter-to-pericenter spans are $0.8$--$62$ Hill radii, which (except for the nearly circular F1 and F4) sweep across the co-orbital region rather than remaining within it.

Consistent with this, no fragment opens a cleared channel over the simulated $1.2$\,kyr; instead the feeding annuli drain smoothly (Section~\ref{sec:accretion}). Over $10^{5}$\,yr, however, even partial gap opening would depress the local surface density below the pool-proportional closure in Equation~(\ref{eq:sigdecay}), so our calculation likely overestimates late-time accretion for any gapped fragment. F3's crossing of the hydrogen-burning boundary is therefore the result most vulnerable to gap opening.

The second caveat is the fixed orbits. The explicit orbital dependence of the growth rate is weak---at fixed mass, $\Sigma\,\Omega R_\mathrm{H}^{2} \propto \Sigma\,a^{1/2}$---but migration changes the surroundings a fragment encounters, which the pool-proportional closure ignores. If maintained at their mean rates, F3's inward drift would carry it into the inner disk, whereas the outward migrators F1 and F4 would move toward the thinning outer edge. The net effect on F3 is therefore ambiguous: inward motion through dense gas could reinforce its dominance, while the scattering and eccentricity pumping in the crowded late disk could instead strand it and cut off its supply. In any case, the measured drifts are episodic and can reverse direction (as for F2), so extrapolating any trend over $10^{5}$\,yr is unrealistic.

Finally, the calculation above deliberately omits mergers, disruption, scattering, and new fragmentation. F6 illustrates this missing dynamical channel: it is being scattered ($e = 0.74$ and rising at the run end), and ejection would produce a free-floating object of planetary mass ($\sim 3$--$4\,M_\mathrm{J}$ including its resident envelope). A single fragmenting disk can therefore yield multiple outcomes---gas giants, brown dwarfs, and (if F6 is ejected) free-floating planets---with a low-mass stellar companion as a fourth, conditional outcome that our calculation attains only if F3 retains its full supply (opening no gap and remaining near its final orbit).

\section{Discussion}\label{sec:discussion}

\subsection{Comparison with Paper I}\label{sec:comparison}

The closest point of contact with \citetalias{Ni_2025ApJ...995...96N} is the initial
fragment mass. Although the two simulations differ in nearly every ingredient
that could plausibly set this mass---the equation of state, the opacity law,
the disk extent, and the surface density profile---the normalized distribution
$m_\mathrm{frag}/M_\mathrm{analytical}$ remains consistent with the
\citetalias{Ni_2025ApJ...995...96N} log-normal
(Section~\ref{sec:initialmass}). This agreement across disparate setups supports the
interpretation of \citetalias{Ni_2025ApJ...995...96N} that the fragmentation scale is
determined locally by the thermodynamic state of the collapsing material,
rather than by the global disk configuration. The subsequent outcomes likewise
match their phenomenology: surviving fragments move to lower
$\alpha_\mathrm{vir}$, while dissolving clumps turn around before reaching
equilibrium (Figure~\ref{fig:census}b), provided that the virial parameter is
evaluated on the scale implied by the adopted equation of state
(Equation~\ref{eq:alphavir}).

The evolution measured here also sharpens the mass question left open by
\citetalias{Ni_2025ApJ...995...96N}. Fragments are born at planetary masses, but
accretion can push them well beyond that scale: by the end of the run the two
most massive objects already exceed the deuterium-burning threshold
($\approx 13\,M_\mathrm{J}$; \citealt{Spiegel_2011ApJ...727...57S}). The conditional
survivor masses in Section~\ref{sec:finalmass} therefore span outcomes from gas
giants to, in the most extreme case, a low-mass stellar companion (if the
dominant accretor retains access to its full gas supply). In this picture, a
GI fragment's final identity---planet, brown dwarf, or low-mass star---is set
less by the fragmentation event itself than by how much of the remaining disk
is ultimately delivered into its Hill sphere.

\subsection{Implications for the evolution of GI fragments}
\label{sec:implications}

Our measurements constrain the assumptions that fragment-evolution models must make about
accretion, migration, spin, and contraction. Across the entire population, the mass-growth rate is
well organized by a single Hill-regulated prescription, with an effective normalization
$C_\mathrm{eff} \approx 4$ and a scatter of a factor of a few; however, the realized growth is
ultimately limited by the rate at which the disk resupplies each Hill sphere. This normalization
matches the two-dimensional result of \citet{Zhu_2012ApJ...746..110Z},
$\dot m = 4\,\Sigma\Omega R_\mathrm{H}^{2}$ (Section~\ref{sec:intro}), and we recover it
independently in three dimensions with radiation transport, suggesting that the Hill-rate
normalization is robust to both. Early-forming fragments in dense environments accrete as fast as
their Hill spheres are replenished, with growth punctuated by events such as mergers, debris
accretion, or arm crossings. By contrast, the latest-forming fragments receive more than an order
of magnitude less fresh gas and grow mainly by binding material already inside their Hill spheres.
Accretion prescriptions that neglect this dependence on environment and formation time will
misestimate the growth of both groups.

Migration is neither uniformly inward nor Type-I-like. Three of seven fragments migrate
outward on average, and the migration rates can exceed linear estimates by up to an order of
magnitude. For two of the three late-forming fragments (F5 and F6), the dominant driver is the
gravity of other clumps rather than disk torques---a few-body process that prescriptions based on
smooth, axisymmetric disks cannot capture. This behavior differs from the results of
\citet{Zhu_2012ApJ...746..110Z}, who reported uniformly inward migration and lost four of thirteen
clumps to tidal disruption. Their runs sample a marginally fragmenting regime (one or two clumps
per disk) that they note avoids the strong clump--clump scattering expected when multiple
fragments coexist, which is precisely the regime our disk occupies. Consistent with this
difference, their disruptions occur after migration to small radii, whereas our fragments remain
between $79$ and $169$\,au, where the stellar tide is weak (Section~\ref{sec:interior}). Other
three-dimensional simulations that evolve multiple fragments simultaneously likewise find migration
that is stochastic in direction \citep{Kubli_2026MNRAS.546ag024K}. As a result, the rapid inward
delivery required for tidal downsizing applies only to a subset of the population here, while F6
illustrates the opposite outcome: clump--clump scattering that can culminate in ejection rather
than orderly inward migration (Section~\ref{sec:migration}).

Competition among fragments also sets the masses they ultimately reach. A single clump with
exclusive access to its annulus can approach an isolation mass of order
$0.1\,M_\odot$, motivating the brown-dwarf expectation of Section~\ref{sec:intro}. When many
fragments draw from the same reservoir, they partition that supply instead: in
Section~\ref{sec:finalmass}, the same Hill-regulated growth law delivers an
isolation-scale mass only to the dominant accretor, while late-forming fragments arrive to find
their annuli already depleted and remain at planetary masses. In this sense, whether GI produces
planets or brown dwarfs is controlled primarily by the multiplicity of fragments sharing the disk,
not by the form of the growth law.

The interiors, finally, are not the non-rotating, purely thermally supported polytropes commonly
assumed as initial conditions for contraction calculations.
Idealized models of fragment interiors usually distinguish a
thermally supported and slowly rotating structure from a rotationally
flattened one \citep[e.g.,][]{Boley_2010Icar..207..509B,
2012MNRAS.427.1725G}. Our fragments
combine the two, with a thermally supported, stratified and convectively
stable core beneath a rotating accretion envelope. Only the merger-processed F1 resembles
the purely thermal case. While the
fragments remain embedded their interiors cannot cool, because radiative
transport falls far short of the accretion power
(Section~\ref{sec:interior}). Contraction calculations for GI-born objects
should therefore start from high-entropy, rotating, stratified states, and
their subsequent evolution will have to shed or reorganize the angular
momentum stored in the envelopes.

One extrapolation follows from the spin measurements. We emphasize that it
is a prediction that this run cannot test. The
six unmerged fragments end the run rotating at $40$--$60$\% of the breakup
rate at their half-mass radii, meaning
$\Omega_\mathrm{spin}(R_\mathrm{half})\,[G(m/2)/R_\mathrm{half}^{3}]^{-1/2}
\approx 0.4$--$0.6$, and are still spinning up by accretion and contraction. If their
specific angular momenta are conserved during further contraction, rotation
becomes limiting at radii of order $0.1$--$1$\,au, where the natural outcome
is the shedding of material into a circumplanetary disk
\citep[e.g.,][]{Boley_2010Icar..207..509B, 2012MNRAS.427.1725G}. On this extrapolation, circumplanetary
disks would be a generic by-product of the contraction of GI fragments,
except where mergers intervene. The single merger in our sample resets the
remnant's spin and tilts it (Section~\ref{sec:spin}), so merger-processed fragments would begin contraction with
little rotation and misaligned spin axes. Testing this prediction requires
following fragments through their second collapse, which the present run
does not reach.

\subsection{Observational signatures for planetary-mass free-floating objects}
\label{sec:FFP}

Our run is too short to decide F6's fate. Its eccentricity climbs steadily to
$e = 0.74$ by the end of the run, driven by its nearest neighbor rather than
by the disk (Section~\ref{sec:migration}), but the encounter is unfinished, and
whether it ends in ejection or in disruption by its neighbor's tide \citep{Calovic_2026MNRAS.545f2097C} depends on the later dynamical evolution of the whole system, including the inner disk that this work does not model. Should it end in ejection, what leaves is the free-floating planetary-mass object Sections~\ref{sec:spin} and
\ref{sec:interior} measured.

Such an object would be distinctive. It leaves while still pre-collapse,
so this channel should place objects that are over-luminous for their mass in
the youngest star-forming regions, still carrying their rotating envelopes. Isolated
planetary-mass objects are indeed observed with dusty disks
\citep{Bayo_2017ApJ...841L..11B, Damian_2025AJ....170..127D}, and one has been
caught in an accretion burst \citep{Almendros_2025ApJ...992L...2A}. A deep survey of Upper Scorpius and
Ophiuchus finds of order a hundred candidate free-floating objects between $4$
and $13\,M_\mathrm{J}$ \citep{MiretRoig_2022NatAs...6...89M}, and microlensing
constrains the free-floating mass function down to Earth masses
\citep{Sumi_2023AJ....166..108S}, with the caveat that a microlensing event
cannot distinguish an unbound object from one on a wide orbit
\citep{Yee_2025AJ....170..132Y}. Our simulation shows that disk fragmentation is a plausible channel for producing them.

\subsection{Caveats}\label{sec:caveats}

There are two numerical caveats we must acknowledge. (i) All results derive from a single realization
of a single disk, so the population statements rest on seven objects and
carry no ensemble variance. (ii) The run covers $3.67$\,kyr, and every evolutionary statement
is censored at that horizon, since the reservoirs are still draining,
F6's scattering is unresolved, and the approach of the youngest
fragments to virial equilibrium is incomplete.

Besides, the following caveats concern the physical model here. (i) The run omits stellar
irradiation and magnetic fields. \citetalias{Ni_2025ApJ...995...96N} discusses why irradiation is subdominant
for the fragmenting outer disk, and non-ideal magnetohydrodynamics is known to alter both
the gravito-turbulent state and clump accretion
\citep[e.g.,][]{Deng_2021NatAs...5..440D, Kubli_2023MNRAS.525.2731K}. (ii) The
fixed $\gamma = 1.4$ equation of state excludes H$_2$ dissociation and the
second collapse it triggers. This remains consistent through the end of the run
(central temperatures stay at or below $\approx 1000$\,K, although at that
temperature vibrational excitation of H$_2$ already begins to soften
$\gamma$) but sets the same
horizon as the duration: the final contraction of the fragments is beyond
this paper. (iii) The simulated disk is also a closed box: nothing
replenishes it from an envelope or an outer disk, and the conditional
masses of Section~\ref{sec:finalmass} divide a fixed budget
accordingly.

\section{Conclusions}\label{sec:conclusions}

We have followed every bound object in a global radiation hydrodynamic simulation
of a fragmenting protoplanetary disk---seven surviving fragments, one disrupted
clump, and one merged clump---over $\approx 1.2$\,kyr with mass and
angular-momentum budgets for each object. Our main conclusions are:

\begin{itemize}

\item Fragments form at $1.4$--$3.0\,M_\mathrm{J}$, and their masses
normalized by the local analytic estimate ($M_\mathrm{analytical} = \Sigma \cdot
\lambda_\mathrm{T}\cdot 2c_s/\Omega$) are consistent with the log-normal
distribution of \citetalias{Ni_2025ApJ...995...96N} despite a different equation of state,
opacity, disk model, and fragmentation environment. The fragmentation
mass scale is therefore set locally, independent of the global disk model.
Survival is not determined by initial mass, as the disrupted and merged
clumps have birth masses comparable to those of the surviving fragments.

\item Growth proceeds in two steps: the delivery of disk gas into each
Hill sphere and the subsequent binding of gas already inside it. A single law,
$\dot m \approx 4\,\Sigma_\mathrm{loc}\Omega R_\mathrm{H}^2$ with a
scatter of a factor of a few, describes all seven fragments. Growth is
limited by tides rather than by pressure ($R_\mathrm{B} > R_\mathrm{H}$),
its geometry is two-dimensional ($R_\mathrm{H} > H$), it is regulated by
feeding zones extending $\approx 2\,R_\mathrm{H}$ on each side of the orbit,
and it is punctuated by three episodes with identifiable mechanisms: a
merger, a debris sweep, and a spiral-arm crossing. The disparity across the
population is governed by delivery: the latest-forming fragments, born in
environments already depleted by the earlier generation, receive more than an
order of magnitude less gas at the same age and grow primarily by consuming the
reservoirs already resident inside their Hill spheres.

\item Migration is bidirectional and rapid, and only one of the seven
fragments drifts at the expected linear Type-I rate and direction. Angular
momentum is exchanged with the disk primarily through the accreted gas itself,
which accounts for nearly all of the change of the total angular momentum for
every fragment except F6. The remaining torque arises from ambient gas for five
fragments, and from the gravity of neighboring clumps for F5 and F6. In
particular, one fragment (F6) is dynamically scattered to $e = 0.74$ by the end
of the run.

\item Accretion establishes the spin. The six unmerged fragments spin up to
$\beta_\mathrm{spin} = 0.09$--$0.15$, rotate differentially with rates declining
outward, and maintain spin axes aligned with their orbits to within $1.5^{\circ}$.
The gas advected by accretion accounts for most of their spin change. In contrast,
the single merger in the sample resets the spin of the remnant, leaving it slowly
rotating and tilted.

\item The interiors are entropy-stratified, largely convectively stable, and cool
inefficiently, with luminosities of only a few percent of the accretion power.
Mass growth proceeds one to two orders of magnitude faster than radiative
cooling, so internal compression is driven by mass addition rather than radiative
energy loss, and compressive $P\,\mathrm{d}V$ work provides the dominant source of
heating. Hydrostatic models integrated from measured entropy profiles generally
reproduce the inner structure of the fragments with no free parameters, while the
accreting envelopes require support beyond gas pressure. This quasi-adiabatic
compression leaves dense cores that are robust against the stellar tide at their
present orbits, and favors hot-start initial conditions for survivors that become
gas giants.

\item Final masses are conditional on the subsequent mass supply and dynamics.
Integrating the growth law forward against the measured gas budget
($202\,M_\mathrm{J}$ in total, of which $140\,M_\mathrm{J}$ is ambient), with the
supply tied to a shared reservoir, the measured end-state efficiencies and
surroundings, fixed orbits, and the whole ambient budget accessible, consumes
the reservoir within $10^{4}$\,yr and partitions it among the fragments
according to their initial accretion rates: F1--F7 reach $15$, $27$, $130$, $15$,
$7.1$, $5.1$, and $2.8\,M_\mathrm{J}$ at $0.1$\,Myr. F2 is a brown dwarf, and
the late fragments F5--F7 remain giant planets in every variant explored. F1 and
F4 land near the deuterium-burning threshold, their variants straddling it, and
F3, which ends as the most massive object in every variant, crosses the
hydrogen-burning boundary under these assumptions. F3's outcome is subject to
two caveats: gap opening, which this super-thermal population would attempt and
which can only lower F3's mass, and rapid inward drift, whose net effect on that
mass remains uncertain. Meanwhile, F6, which is undergoing scattering, may be
ejected to join the free-floating planet population at its present $3$ to
$4\,M_\mathrm{J}$, resident envelope included.

\end{itemize}

Natural next steps include simulations that reach the second collapse of the
fragments, including $\mathrm{H}_2$ dissociation, longer durations, and the
magnetic fields and stellar irradiation omitted here. Such calculations will
test the prediction that contraction converts the measured spins into
circumplanetary disks, and determine where the termination of accretion leaves
the final mass function.

\begin{acknowledgments}

\end{acknowledgments}

\bibliography{sample701}{}

@ARTICLE{Hopkins_2015MNRAS.450...53H,
       author = {{Hopkins}, Philip F.},
        title = "{A new class of accurate, mesh-free hydrodynamic simulation methods}",
      journal = {\mnras},
         year = 2015,
        month = jun,
       volume = {450},
       number = {1},
        pages = {53-110},
          doi = {10.1093/mnras/stv195},
archivePrefix = {arXiv},
       eprint = {1409.7395},
 primaryClass = {astro-ph.CO},
       adsurl = {https://ui.adsabs.harvard.edu/abs/2015MNRAS.450...53H}
}

@ARTICLE{Levermore_1984JQSRT..31..149L,
       author = {{Levermore}, C.~D.},
        title = "{Relating Eddington factors to flux limiters.}",
      journal = {\jqsrt},
         year = 1984,
        month = feb,
       volume = {31},
       number = {2},
        pages = {149-160},
          doi = {10.1016/0022-4073(84)90112-2},
       adsurl = {https://ui.adsabs.harvard.edu/abs/1984JQSRT..31..149L}
}

@ARTICLE{Ni_2025A&A...699A.282N,
       author = {{Ni}, Yang and {Li}, Hui and {Vogelsberger}, Mark and {Sales}, Laura V. and {Marinacci}, Federico and {Torrey}, Paul},
        title = "{The life cycle of giant molecular clouds in simulated Milky Way-mass galaxies}",
      journal = {\aap},
         year = 2025,
        month = jul,
       volume = {699},
          eid = {A282},
        pages = {A282},
          doi = {10.1051/0004-6361/202554126},
archivePrefix = {arXiv},
       eprint = {2502.12256},
 primaryClass = {astro-ph.GA},
       adsurl = {https://ui.adsabs.harvard.edu/abs/2025A&A...699A.282N}
}

@ARTICLE{Ni_2025ApJ...995...96N,
       author = {{Ni}, Yang and {Deng}, Hongping and {Bai}, Xue-Ning},
        title = "{Radiation Hydrodynamics of Self-gravitating Protoplanetary Disks. I. Direct Formation of Gas Giants via Disk Fragmentation}",
      journal = {\apj},
         year = 2025,
        month = dec,
       volume = {995},
       number = {1},
          eid = {96},
        pages = {96},
          doi = {10.3847/1538-4357/ae16a4},
archivePrefix = {arXiv},
       eprint = {2510.19915},
 primaryClass = {astro-ph.EP},
       adsurl = {https://ui.adsabs.harvard.edu/abs/2025ApJ...995...96N}
}

@ARTICLE{Zhu_2021MNRAS.508..453Z,
       author = {{Zhu}, Zhaohuan and {Jiang}, Yan-Fei and {Baehr}, Hans and {Youdin}, Andrew N. and {Armitage}, Philip J. and {Martin}, Rebecca G.},
        title = "{Global 3D radiation hydrodynamic simulations of proto-Jupiter's convective envelope}",
      journal = {\mnras},
         year = 2021,
        month = nov,
       volume = {508},
       number = {1},
        pages = {453-474},
          doi = {10.1093/mnras/stab2517},
archivePrefix = {arXiv},
       eprint = {2106.12003},
 primaryClass = {astro-ph.EP},
       adsurl = {https://ui.adsabs.harvard.edu/abs/2021MNRAS.508..453Z}
}

@ARTICLE{2002ApJ...565.1257T,
       author = {{Tanaka}, Hidekazu and {Takeuchi}, Taku and {Ward}, William R.},
        title = "{Three-Dimensional Interaction between a Planet and an Isothermal Gaseous Disk. I. Corotation and Lindblad Torques and Planet Migration}",
      journal = {\apj},
         year = 2002,
        month = feb,
       volume = {565},
       number = {2},
        pages = {1257-1274},
          doi = {10.1086/324713},
       adsurl = {https://ui.adsabs.harvard.edu/abs/2002ApJ...565.1257T}
}

@ARTICLE{Spiegel_2011ApJ...727...57S,
       author = {{Spiegel}, David S. and {Burrows}, Adam and {Milsom}, John A.},
        title = "{The Deuterium-burning Mass Limit for Brown Dwarfs and Giant Planets}",
      journal = {\apj},
         year = 2011,
        month = jan,
       volume = {727},
       number = {1},
          eid = {57},
        pages = {57},
          doi = {10.1088/0004-637X/727/1/57},
archivePrefix = {arXiv},
       eprint = {1008.5150},
 primaryClass = {astro-ph.EP},
       adsurl = {https://ui.adsabs.harvard.edu/abs/2011ApJ...727...57S}
}

@ARTICLE{2012MNRAS.427.1725G,
       author = {{Galvagni}, M. and {Hayfield}, T. and {Boley}, A. and {Mayer}, L. and {Ro{\v{s}}kar}, R. and {Saha}, P.},
        title = "{The collapse of protoplanetary clumps formed through disc instability: 3D simulations of the pre-dissociation phase}",
      journal = {\mnras},
         year = 2012,
        month = dec,
       volume = {427},
       number = {2},
        pages = {1725-1740},
          doi = {10.1111/j.1365-2966.2012.22096.x},
archivePrefix = {arXiv},
       eprint = {1209.2129},
 primaryClass = {astro-ph.EP},
       adsurl = {https://ui.adsabs.harvard.edu/abs/2012MNRAS.427.1725G}
}

@ARTICLE{Deng_2021NatAs...5..440D,
       author = {{Deng}, Hongping and {Mayer}, Lucio and {Helled}, Ravit},
        title = "{Formation of intermediate-mass planets via magnetically controlled disk fragmentation}",
      journal = {Nature Astronomy},
         year = 2021,
        month = feb,
       volume = {5},
        pages = {440-444},
          doi = {10.1038/s41550-020-01297-6},
archivePrefix = {arXiv},
       eprint = {2101.01331},
 primaryClass = {astro-ph.EP},
       adsurl = {https://ui.adsabs.harvard.edu/abs/2021NatAs...5..440D}
}

@ARTICLE{Kubli_2023MNRAS.525.2731K,
       author = {{Kubli}, Noah and {Mayer}, Lucio and {Deng}, Hongping},
        title = "{Characterizing fragmentation and sub-Jovian clump properties in magnetized young protoplanetary discs}",
      journal = {\mnras},
         year = 2023,
        month = oct,
       volume = {525},
       number = {2},
        pages = {2731-2749},
          doi = {10.1093/mnras/stad2478},
archivePrefix = {arXiv},
       eprint = {2303.04163},
 primaryClass = {astro-ph.EP},
       adsurl = {https://ui.adsabs.harvard.edu/abs/2023MNRAS.525.2731K}
}

@ARTICLE{Boley_2010Icar..207..509B,
       author = {{Boley}, Aaron C. and {Hayfield}, Tristen and {Mayer}, Lucio and {Durisen}, Richard H.},
        title = "{Clumps in the outer disk by disk instability: Why they are initially gas giants and the legacy of disruption}",
      journal = {\icarus},
         year = 2010,
        month = jun,
       volume = {207},
       number = {2},
        pages = {509-516},
          doi = {10.1016/j.icarus.2010.01.015},
archivePrefix = {arXiv},
       eprint = {0909.4543},
 primaryClass = {astro-ph.EP},
       adsurl = {https://ui.adsabs.harvard.edu/abs/2010Icar..207..509B}
}

@ARTICLE{Pollack_1996Icar..124...62P,
       author = {{Pollack}, James B. and {Hubickyj}, Olenka and {Bodenheimer}, Peter and {Lissauer}, Jack J. and {Podolak}, Morris and {Greenzweig}, Yuval},
        title = "{Formation of the Giant Planets by Concurrent Accretion of Solids and Gas}",
      journal = {\icarus},
         year = 1996,
        month = nov,
       volume = {124},
       number = {1},
        pages = {62-85},
          doi = {10.1006/icar.1996.0190},
       adsurl = {https://ui.adsabs.harvard.edu/abs/1996Icar..124...62P}
}

@ARTICLE{Lambrechts_2012A&A...544A..32L,
       author = {{Lambrechts}, M. and {Johansen}, A.},
        title = "{Rapid growth of gas-giant cores by pebble accretion}",
      journal = {\aap},
         year = 2012,
        month = aug,
       volume = {544},
          eid = {A32},
        pages = {A32},
          doi = {10.1051/0004-6361/201219127},
archivePrefix = {arXiv},
       eprint = {1205.3030},
 primaryClass = {astro-ph.EP},
       adsurl = {https://ui.adsabs.harvard.edu/abs/2012A&A...544A..32L}
}

@ARTICLE{Cameron_1978M&P....18....5C,
       author = {{Cameron}, A.~G.~W.},
        title = "{Physics of the Primitive Solar Accretion Disk}",
      journal = {Moon and Planets},
         year = 1978,
        month = feb,
       volume = {18},
       number = {1},
        pages = {5-40},
          doi = {10.1007/BF00896696},
       adsurl = {https://ui.adsabs.harvard.edu/abs/1978M&P....18....5C}
}

@ARTICLE{Boss_1997Sci...276.1836B,
       author = {{Boss}, A.~P.},
        title = "{Giant planet formation by gravitational instability.}",
      journal = {Science},
         year = 1997,
        month = jan,
       volume = {276},
        pages = {1836-1839},
          doi = {10.1126/science.276.5320.1836},
       adsurl = {https://ui.adsabs.harvard.edu/abs/1997Sci...276.1836B}
}

@ARTICLE{Kratter_2016ARA&A..54..271K,
       author = {{Kratter}, Kaitlin and {Lodato}, Giuseppe},
        title = "{Gravitational Instabilities in Circumstellar Disks}",
      journal = {\araa},
         year = 2016,
        month = sep,
       volume = {54},
        pages = {271-311},
          doi = {10.1146/annurev-astro-081915-023307},
archivePrefix = {arXiv},
       eprint = {1603.01280},
 primaryClass = {astro-ph.SR},
       adsurl = {https://ui.adsabs.harvard.edu/abs/2016ARA&A..54..271K}
}

@ARTICLE{Forgan2011jeans,
       author = {{Forgan}, Duncan and {Rice}, Ken},
        title = "{The Jeans mass as a fundamental measure of self-gravitating disc fragmentation and initial fragment mass}",
      journal = {\mnras},
         year = 2011,
        month = nov,
       volume = {417},
       number = {3},
        pages = {1928-1937},
          doi = {10.1111/j.1365-2966.2011.19380.x},
archivePrefix = {arXiv},
       eprint = {1107.0831},
 primaryClass = {astro-ph.EP},
       adsurl = {https://ui.adsabs.harvard.edu/abs/2011MNRAS.417.1928F}
}

@ARTICLE{Xu_2025ApJ...986...91X,
       author = {{Xu}, Wenrui and {Jiang}, Yan-Fei and {Kunz}, Matthew W. and {Stone}, James M.},
        title = "{Global Simulations of Gravitational Instability in Protostellar Disks with Full Radiation Transport. I. Stochastic Fragmentation with Optical-depth-dependent Rate and Universal Fragment Mass}",
      journal = {\apj},
         year = 2025,
        month = jun,
       volume = {986},
       number = {1},
          eid = {91},
        pages = {91},
          doi = {10.3847/1538-4357/add14a},
archivePrefix = {arXiv},
       eprint = {2410.12042},
 primaryClass = {astro-ph.EP},
       adsurl = {https://ui.adsabs.harvard.edu/abs/2025ApJ...986...91X}
}

@ARTICLE{Marleau_2014MNRAS.437.1378M,
       author = {{Marleau}, G.-D. and {Cumming}, A.},
        title = "{Constraining the initial entropy of directly detected exoplanets}",
      journal = {\mnras},
         year = 2014,
        month = jan,
       volume = {437},
       number = {2},
        pages = {1378-1399},
          doi = {10.1093/mnras/stt1967},
archivePrefix = {arXiv},
       eprint = {1302.1517},
 primaryClass = {astro-ph.EP},
       adsurl = {https://ui.adsabs.harvard.edu/abs/2014MNRAS.437.1378M}
}

@ARTICLE{Burrows_1997ApJ...491..856B,
       author = {{Burrows}, A. and {Marley}, M. and {Hubbard}, W.~B. and {Lunine}, J.~I. and {Guillot}, T. and {Saumon}, D. and {Freedman}, R. and {Sudarsky}, D. and {Sharp}, C.},
        title = "{A Nongray Theory of Extrasolar Giant Planets and Brown Dwarfs}",
      journal = {\apj},
         year = 1997,
        month = dec,
       volume = {491},
       number = {2},
        pages = {856-875},
          doi = {10.1086/305002},
archivePrefix = {arXiv},
       eprint = {astro-ph/9705201},
 primaryClass = {astro-ph},
       adsurl = {https://ui.adsabs.harvard.edu/abs/1997ApJ...491..856B}
}

@ARTICLE{Marley_2007ApJ...655..541M,
       author = {{Marley}, Mark S. and {Fortney}, Jonathan J. and {Hubickyj}, Olenka and {Bodenheimer}, Peter and {Lissauer}, Jack J.},
        title = "{On the Luminosity of Young Jupiters}",
      journal = {\apj},
         year = 2007,
        month = jan,
       volume = {655},
       number = {1},
        pages = {541-549},
          doi = {10.1086/509759},
archivePrefix = {arXiv},
       eprint = {astro-ph/0609739},
 primaryClass = {astro-ph},
       adsurl = {https://ui.adsabs.harvard.edu/abs/2007ApJ...655..541M}
}

@ARTICLE{Spiegel_2012ApJ...745..174S,
       author = {{Spiegel}, David S. and {Burrows}, Adam},
        title = "{Spectral and Photometric Diagnostics of Giant Planet Formation Scenarios}",
      journal = {\apj},
         year = 2012,
        month = feb,
       volume = {745},
       number = {2},
          eid = {174},
        pages = {174},
          doi = {10.1088/0004-637X/745/2/174},
archivePrefix = {arXiv},
       eprint = {1108.5172},
 primaryClass = {astro-ph.EP},
       adsurl = {https://ui.adsabs.harvard.edu/abs/2012ApJ...745..174S}
}

@ARTICLE{Crida_2006Icar..181..587C,
       author = {{Crida}, A. and {Morbidelli}, A. and {Masset}, F.},
        title = "{On the width and shape of gaps in protoplanetary disks}",
      journal = {\icarus},
         year = 2006,
        month = apr,
       volume = {181},
       number = {2},
        pages = {587-604},
          doi = {10.1016/j.icarus.2005.10.007},
archivePrefix = {arXiv},
       eprint = {astro-ph/0511082},
 primaryClass = {astro-ph},
       adsurl = {https://ui.adsabs.harvard.edu/abs/2006Icar..181..587C}
}

@ARTICLE{Calovic_2026MNRAS.545f2097C,
       author = {{{\'C}alovi{\'c}}, Aleksandra and {Nayakshin}, Sergei and {Casewell}, Sarah and {Miret-Roig}, N{\'u}ria},
        title = "{Disc fragmentation -- I. Ejection of Jupiter-mass free floating planets from growing binary systems}",
      journal = {\mnras},
         year = 2026,
        month = jan,
       volume = {545},
       number = {3},
          eid = {staf2097},
        pages = {staf2097},
          doi = {10.1093/mnras/staf2097},
archivePrefix = {arXiv},
       eprint = {2511.16508},
 primaryClass = {astro-ph.EP},
       adsurl = {https://ui.adsabs.harvard.edu/abs/2026MNRAS.545f2097C}
}

@ARTICLE{MiretRoig_2022NatAs...6...89M,
       author = {{Miret-Roig}, N{\'u}ria and {Bouy}, Herv{\'e} and {Raymond}, Sean N. and {Tamura}, Motohide and {Bertin}, Emmanuel and {Barrado}, David and {Olivares}, Javier and {Galli}, Phillip A.~B. and {Cuillandre}, Jean-Charles and {Sarro}, Luis Manuel and {Berihuete}, Angel and {Hu{\'e}lamo}, Nuria},
        title = "{A rich population of free-floating planets in the Upper Scorpius young stellar association}",
      journal = {Nature Astronomy},
         year = 2022,
        month = feb,
       volume = {6},
        pages = {89-97},
          doi = {10.1038/s41550-021-01513-x},
archivePrefix = {arXiv},
       eprint = {2112.11999},
 primaryClass = {astro-ph.EP},
       adsurl = {https://ui.adsabs.harvard.edu/abs/2022NatAs...6...89M}
}

@ARTICLE{Sumi_2023AJ....166..108S,
       author = {{Sumi}, Takahiro and {Koshimoto}, Naoki and {Bennett}, David P. and {Rattenbury}, Nicholas J. and {Abe}, Fumio and {Barry}, Richard and {Bhattacharya}, Aparna and {Bond}, Ian A. and {Fujii}, Hirosane and {Fukui}, Akihiko and {Hamada}, Ryusei and {Hirao}, Yuki and {Silva}, Stela Ishitani and {Itow}, Yoshitaka and {Kirikawa}, Rintaro and {Kondo}, Iona and {Matsubara}, Yutaka and {Miyazaki}, Shota and {Muraki}, Yasushi and {Olmschenk}, Greg and {Ranc}, Cl{\'e}ment and {Satoh}, Yuki and {Suzuki}, Daisuke and {Tomoyoshi}, Mio and {Tristram}, Paul. J. and {Vandorou}, Aikaterini and {Yama}, Hibiki and {Yamashita}, Kansuke},
        title = "{Free-floating Planet Mass Function from MOA-II 9 yr Survey toward the Galactic Bulge}",
      journal = {\aj},
         year = 2023,
        month = sep,
       volume = {166},
       number = {3},
          eid = {108},
        pages = {108},
          doi = {10.3847/1538-3881/ace688},
archivePrefix = {arXiv},
       eprint = {2303.08280},
 primaryClass = {astro-ph.EP},
       adsurl = {https://ui.adsabs.harvard.edu/abs/2023AJ....166..108S}
}

@ARTICLE{Yee_2025AJ....170..132Y,
       author = {{Yee}, Jennifer C. and {Kenyon}, Scott J.},
        title = "{Microlensing Constraints on the Stellar and Planetary Mass Functions}",
      journal = {\aj},
         year = 2025,
        month = aug,
       volume = {170},
       number = {2},
          eid = {132},
        pages = {132},
          doi = {10.3847/1538-3881/adeb84},
archivePrefix = {arXiv},
       eprint = {2503.11597},
 primaryClass = {astro-ph.EP},
       adsurl = {https://ui.adsabs.harvard.edu/abs/2025AJ....170..132Y}
}

@ARTICLE{Bayo_2017ApJ...841L..11B,
       author = {{Bayo}, Amelia and {Joergens}, Viki and {Liu}, Yao and {Brauer}, Robert and {Olofsson}, Johan and {Arancibia}, Javier and {Pinilla}, Paola and {Wolf}, Sebastian and {Ruge}, Jan Philipp and {Henning}, Thomas and {Natta}, Antonella and {Johnston}, Katharine G. and {Bonnefoy}, Mickael and {Beuther}, Henrik and {Chauvin}, Gael},
        title = "{First Millimeter Detection of the Disk around a Young, Isolated, Planetary-mass Object}",
      journal = {\apjl},
         year = 2017,
        month = may,
       volume = {841},
       number = {1},
          eid = {L11},
        pages = {L11},
          doi = {10.3847/2041-8213/aa7046},
archivePrefix = {arXiv},
       eprint = {1705.06378},
 primaryClass = {astro-ph.SR},
       adsurl = {https://ui.adsabs.harvard.edu/abs/2017ApJ...841L..11B}
}

@ARTICLE{Damian_2025AJ....170..127D,
       author = {{Damian}, Belinda and {Scholz}, Aleks and {Jayawardhana}, Ray and {Almendros-Abad}, V. and {Flagg}, Laura and {Mu{\v{z}}i{\'c}}, Koraljka and {Natta}, Antonella and {Pinilla}, Paola and {Testi}, Leonardo},
        title = "{Spectroscopy of Free-floating Planetary-mass Objects and Their Disks with JWST}",
      journal = {\aj},
         year = 2025,
        month = aug,
       volume = {170},
       number = {2},
          eid = {127},
        pages = {127},
          doi = {10.3847/1538-3881/adea50},
archivePrefix = {arXiv},
       eprint = {2507.05155},
 primaryClass = {astro-ph.EP},
       adsurl = {https://ui.adsabs.harvard.edu/abs/2025AJ....170..127D}
}

@ARTICLE{Almendros_2025ApJ...992L...2A,
       author = {{Almendros-Abad}, Victor and {Scholz}, Aleks and {Damian}, Belinda and {Jayawardhana}, Ray and {Bayo}, Amelia and {Flagg}, Laura and {Mu{\v{z}}i{\'c}}, Koraljka and {Natta}, Antonella and {Pinilla}, Paola and {Testi}, Leonardo},
        title = "{Discovery of an Accretion Burst in a Free-floating Planetary-mass Object}",
      journal = {\apjl},
         year = 2025,
        month = oct,
       volume = {992},
       number = {1},
          eid = {L2},
        pages = {L2},
          doi = {10.3847/2041-8213/ae09a8},
archivePrefix = {arXiv},
       eprint = {2510.01747},
 primaryClass = {astro-ph.SR},
       adsurl = {https://ui.adsabs.harvard.edu/abs/2025ApJ...992L...2A}
}

@ARTICLE{Zhu_2012ApJ...746..110Z,
       author = {{Zhu}, Zhaohuan and {Hartmann}, Lee and {Nelson}, Richard P. and {Gammie}, Charles F.},
        title = "{Challenges in Forming Planets by Gravitational Instability: Disk Irradiation and Clump Migration, Accretion, and Tidal Destruction}",
      journal = {\apj},
         year = 2012,
        month = feb,
       volume = {746},
       number = {1},
          eid = {110},
        pages = {110},
          doi = {10.1088/0004-637X/746/1/110},
archivePrefix = {arXiv},
       eprint = {1111.6943},
 primaryClass = {astro-ph.SR},
       adsurl = {https://ui.adsabs.harvard.edu/abs/2012ApJ...746..110Z}
}

@ARTICLE{Tsukamoto_2013MNRAS.436.1667T,
       author = {{Tsukamoto}, Yusuke and {Machida}, Masahiro N. and {Inutsuka}, Shu-ichiro},
        title = "{Formation, orbital and thermal evolution, and survival of planetary-mass clumps in the early phase of circumstellar disc evolution}",
      journal = {\mnras},
         year = 2013,
        month = dec,
       volume = {436},
       number = {2},
        pages = {1667-1673},
          doi = {10.1093/mnras/stt1684},
archivePrefix = {arXiv},
       eprint = {1307.6910},
 primaryClass = {astro-ph.SR},
       adsurl = {https://ui.adsabs.harvard.edu/abs/2013MNRAS.436.1667T}
}

@ARTICLE{Baruteau_2011MNRAS.416.1971B,
       author = {{Baruteau}, Cl{\'e}ment and {Meru}, Farzana and {Paardekooper}, Sijme-Jan},
        title = "{Rapid inward migration of planets formed by gravitational instability}",
      journal = {\mnras},
         year = 2011,
        month = sep,
       volume = {416},
       number = {3},
        pages = {1971-1982},
          doi = {10.1111/j.1365-2966.2011.19172.x},
archivePrefix = {arXiv},
       eprint = {1106.0487},
 primaryClass = {astro-ph.EP},
       adsurl = {https://ui.adsabs.harvard.edu/abs/2011MNRAS.416.1971B}
}

@ARTICLE{Stamatellos_2015ApJ...810L..11S,
       author = {{Stamatellos}, Dimitris},
        title = "{The Migration of Gas Giant Planets in Gravitationally Unstable Disks}",
      journal = {\apjl},
         year = 2015,
        month = sep,
       volume = {810},
       number = {1},
          eid = {L11},
        pages = {L11},
          doi = {10.1088/2041-8205/810/1/L11},
archivePrefix = {arXiv},
       eprint = {1508.01196},
 primaryClass = {astro-ph.EP},
       adsurl = {https://ui.adsabs.harvard.edu/abs/2015ApJ...810L..11S}
}

@ARTICLE{Kubli_2026MNRAS.546ag024K,
       author = {{Kubli}, Noah and {Mayer}, Lucio and {Deng}, Hongping and {Lin}, Douglas N.~C.},
        title = "{The stochastic nature of migration of disc instability protoplanets in three-dimensional hydrodynamical and MHD simulations of fragmenting discs}",
      journal = {\mnras},
         year = 2026,
        month = mar,
       volume = {546},
       number = {3},
          eid = {stag024},
        pages = {stag024},
          doi = {10.1093/mnras/stag024},
archivePrefix = {arXiv},
       eprint = {2503.01973},
 primaryClass = {astro-ph.EP},
       adsurl = {https://ui.adsabs.harvard.edu/abs/2026MNRAS.546ag024K}
}

@ARTICLE{Forgan_2018MNRAS.474.5036F,
       author = {{Forgan}, D.~H. and {Hall}, C. and {Meru}, F. and {Rice}, W.~K.~M.},
        title = "{Towards a population synthesis model of self-gravitating disc fragmentation and tidal downsizing II: the effect of fragment-fragment interactions}",
      journal = {\mnras},
         year = 2018,
        month = mar,
       volume = {474},
       number = {4},
        pages = {5036-5048},
          doi = {10.1093/mnras/stx2870},
archivePrefix = {arXiv},
       eprint = {1711.01133},
 primaryClass = {astro-ph.EP},
       adsurl = {https://ui.adsabs.harvard.edu/abs/2018MNRAS.474.5036F}
}

@ARTICLE{Schib_2025A&A...704A..28S,
       author = {{Schib}, O. and {Mordasini}, C. and {Emsenhuber}, A. and {Helled}, R.},
        title = "{DIPSY: A new Disc Instability Population SYnthesis: II. The Populations of Companions Formed Through Disc Instability}",
      journal = {\aap},
         year = 2025,
        month = dec,
       volume = {704},
          eid = {A28},
        pages = {A28},
          doi = {10.1051/0004-6361/202556261},
archivePrefix = {arXiv},
       eprint = {2510.02437},
 primaryClass = {astro-ph.EP},
       adsurl = {https://ui.adsabs.harvard.edu/abs/2025A&A...704A..28S}
}

@ARTICLE{Fletcher_2019MNRAS.486.4398F,
       author = {{Fletcher}, M. and {Nayakshin}, S. and {Stamatellos}, D. and {Dehnen}, W. and {Meru}, F. and {Mayer}, L. and {Deng}, H. and {Rice}, K.},
        title = "{Giant planets and brown dwarfs on wide orbits: a code comparison project}",
      journal = {\mnras},
         year = 2019,
        month = jul,
       volume = {486},
       number = {3},
        pages = {4398-4413},
          doi = {10.1093/mnras/stz1123},
archivePrefix = {arXiv},
       eprint = {1901.08089},
 primaryClass = {astro-ph.EP},
       adsurl = {https://ui.adsabs.harvard.edu/abs/2019MNRAS.486.4398F}
}

@ARTICLE{Currie_2022NatAs...6..751C,
       author = {{Currie}, Thayne and {Lawson}, Kellen and {Schneider}, Glenn and {Lyra}, Wladimir and {Wisniewski}, John and {Grady}, Carol and {Guyon}, Olivier and {Tamura}, Motohide and {Kotani}, Takayuki and {Kawahara}, Hajime and {Brandt}, Timothy and {Uyama}, Taichi and {Muto}, Takayuki and {Dong}, Ruobing and {Kudo}, Tomoyuki and {Hashimoto}, Jun and {Fukagawa}, Misato and {Wagner}, Kevin and {Lozi}, Julien and {Chilcote}, Jeffrey and {Tobin}, Taylor and {Groff}, Tyler and {Ward-Duong}, Kimberly and {Januszewski}, William and {Norris}, Barnaby and {Tuthill}, Peter and {van der Marel}, Nienke and {Sitko}, Michael and {Deo}, Vincent and {Vievard}, Sebastien and {Jovanovic}, Nemanja and {Martinache}, Frantz and {Skaf}, Nour},
        title = "{Images of embedded Jovian planet formation at a wide separation around AB Aurigae}",
      journal = {Nature Astronomy},
         year = 2022,
        month = apr,
       volume = {6},
        pages = {751-759},
          doi = {10.1038/s41550-022-01634-x},
archivePrefix = {arXiv},
       eprint = {2204.00633},
 primaryClass = {astro-ph.EP},
       adsurl = {https://ui.adsabs.harvard.edu/abs/2022NatAs...6..751C}
}

@ARTICLE{Speedie_2024Natur.633...58S,
       author = {{Speedie}, Jessica and {Dong}, Ruobing and {Hall}, Cassandra and {Longarini}, Cristiano and {Veronesi}, Benedetta and {Paneque-Carre{\~n}o}, Teresa and {Lodato}, Giuseppe and {Tang}, Ya-Wen and {Teague}, Richard and {Hashimoto}, Jun},
        title = "{Gravitational instability in a planet-forming disk}",
      journal = {\nat},
         year = 2024,
        month = sep,
       volume = {633},
       number = {8028},
        pages = {58-62},
          doi = {10.1038/s41586-024-07877-0},
archivePrefix = {arXiv},
       eprint = {2409.02196},
 primaryClass = {astro-ph.EP},
       adsurl = {https://ui.adsabs.harvard.edu/abs/2024Natur.633...58S}
}

@ARTICLE{Kanno_2013PASJ...65...72K,
       author = {{Kanno}, Yuji and {Harada}, Tetsuya and {Hanawa}, Tomoyuki},
        title = "{Kinetic Scheme for Solving the M1 Model of Radiative Transfer}",
      journal = {\pasj},
         year = 2013,
        month = aug,
       volume = {65},
       number = {4},
          eid = {72},
        pages = {72},
          doi = {10.1093/pasj/65.4.72},
archivePrefix = {arXiv},
       eprint = {1303.6805},
 primaryClass = {astro-ph.SR},
       adsurl = {https://ui.adsabs.harvard.edu/abs/2013PASJ...65...72K}
}

@ARTICLE{Deng_2017ApJ...847...43D,
       author = {{Deng}, Hongping and {Mayer}, Lucio and {Meru}, Farzana},
        title = "{Convergence of the Critical Cooling Rate for Protoplanetary Disk Fragmentation Achieved: The Key Role of Numerical Dissipation of Angular Momentum}",
      journal = {\apj},
         year = 2017,
        month = sep,
       volume = {847},
       number = {1},
          eid = {43},
        pages = {43},
          doi = {10.3847/1538-4357/aa872b},
archivePrefix = {arXiv},
       eprint = {1706.00417},
 primaryClass = {astro-ph.EP},
       adsurl = {https://ui.adsabs.harvard.edu/abs/2017ApJ...847...43D}
}

@ARTICLE{Machida_2010MNRAS.405.1227M,
       author = {{Machida}, Masahiro N. and {Kokubo}, Eiichiro and {Inutsuka}, Shu-Ichiro and {Matsumoto}, Tomoaki},
        title = "{Gas accretion onto a protoplanet and formation of a gas giant planet}",
      journal = {\mnras},
         year = 2010,
        month = jun,
       volume = {405},
       number = {2},
        pages = {1227-1243},
          doi = {10.1111/j.1365-2966.2010.16527.x},
archivePrefix = {arXiv},
       eprint = {1002.3002},
 primaryClass = {astro-ph.SR},
       adsurl = {https://ui.adsabs.harvard.edu/abs/2010MNRAS.405.1227M}
}

@ARTICLE{Kuwahara_2019A&A...623A.179K,
       author = {{Kuwahara}, Ayumu and {Kurokawa}, Hiroyuki and {Ida}, Shigeru},
        title = "{Gas flow around a planet embedded in a protoplanetary disc. Dependence on planetary mass}",
      journal = {\aap},
         year = 2019,
        month = mar,
       volume = {623},
          eid = {A179},
        pages = {A179},
          doi = {10.1051/0004-6361/201833997},
archivePrefix = {arXiv},
       eprint = {1901.08253},
 primaryClass = {astro-ph.EP},
       adsurl = {https://ui.adsabs.harvard.edu/abs/2019A&A...623A.179K}
}

@ARTICLE{Bethune_2019MNRAS.488.2365B,
       author = {{B{\'e}thune}, William and {Rafikov}, Roman R.},
        title = "{Envelopes of embedded super-Earths - II. Three-dimensional isothermal simulations}",
      journal = {\mnras},
         year = 2019,
        month = sep,
       volume = {488},
       number = {2},
        pages = {2365-2379},
          doi = {10.1093/mnras/stz1870},
archivePrefix = {arXiv},
       eprint = {1907.02763},
 primaryClass = {astro-ph.EP},
       adsurl = {https://ui.adsabs.harvard.edu/abs/2019MNRAS.488.2365B}
}
\bibliographystyle{aasjournalv7}
\end{CJK*}
\end{document}